\documentclass{aastex63}
\usepackage{amsmath}
\usepackage{multirow}
\usepackage{natbib}
\usepackage{color}

\submitjournal{ApJ}

\shortauthors{Manek \& Brummell}

\begin{document}

\title{On the origin of the solar hemispherical helicity rules: Simulations of the rise of magnetic flux concentrations in a background field}

\correspondingauthor{Bhishek Manek}
\email{bmanek@ucsc.edu}

\author[0000-0002-2244-5436]{Bhishek Manek}
\author[0000-0003-4350-5183]{Nicholas Brummell}

\affiliation{Department of Applied Mathematics, Jack Baskin School of Engineering, University of California Santa Cruz, \\ 1156 High Street, Santa Cruz, California 95064, USA}

\begin{abstract}

Solar active regions and sunspots are believed to be formed by the emergence of strong toroidal magnetic flux from the solar interior. Modeling of such events has focused on the dynamics of compact magnetic entities, colloquially known as ``flux tubes", often considered to be isolated magnetic structures embedded in an otherwise field-free environment. In this paper, we show that relaxing such idealized assumptions can lead to surprisingly different dynamics. We consider the rise of tube-like \textit{flux concentrations} embedded in a large-scale volume-filling horizontal field in an initially quiescent adiabatically-stratified compressible fluid.  In a previous letter, we revealed the unexpected major result that concentrations that have their twist aligned with the background field at the bottom of the tube are more likely to rise than the opposite orientation (for certain values of the parameters).  This bias leads to a selection rule which, when applied to solar dynamics, is in agreement with the observations known as the solar hemispheric helicity rule(s) (SHHR).  Here, we examine this selection mechanism in more detail than was possible in the earlier letter.  We explore the dependence on the parameters via simulations, delineating the Selective Rise Regime (SRR), where the bias operates.  We provide a theoretical model to predict and explain the simulation dynamics. Furthermore, we create synthetic helicity maps from Monte Carlo simulations to mimic the SHHR observations  and demonstrate that our mechanism explains the observed scatter in the rule and its variation over the solar cycle. 

\end{abstract}


\section{Introduction} 
\label{sec:intro}

Large-scale solar surface magnetic fields, such as sunspots embedded in active regions, are widely believed to be manifestations of deep interior magnetic field rising to the surface due to magnetic buoyancy \citep{Parker:Buoyancy:1975}. While small-scale fields (on the scale of observable solar surface velocities) exhibit very turbulent behaviour, the large-scale fields appear to have surprisingly ordered dynamics. Observational studies have clearly established the cyclic nature of such large-scale solar magnetic fields, often exhibited as the butterfly pattern of the emergence of active regions and surface sunspots.  Further regularity rules are Hale's Law, associated with active region polarity, and Joy's Law, associated with their longitudinal tilt  \citep{Hale:Ellerman:Ferdinand:Nicholson:1919}.

More recent observations of the solar surface have focused on other structural properties of the active regions. Magnetic helicity \citep{Moffatt:1969}, defined as $H_{M}=\int_{V}{\bf A}\cdot (\nabla \times {\bf A}) ~dV$ (where ${\bf B}= \nabla \times {\bf A}$ is the magnetic field, ${\bf A}$ is its potential, and $V$ is a volume), measures the complexity of the active regions in terms of the twisting, kinking and linking of magnetic field lines there. Magnetic helicity is an important dynamical quantity for a number of reasons.  For example, in ideal MHD, $H_{M}$ is conserved and is, therefore, a significant constraint for dynamo theory \citep{Berger:1984, Berger:Field:1984, Blackman:Field:2002}.  Furthermore,  the release of energy stored in helical fields by reconnection (requiring some resistivity) in the solar atmosphere is thought to be a driver of energetic events, such as flares, jets, and coronal mass ejections (CMEs) \citep[e.g.][]{Low:1996, Amari:Luciani:Aly:Mikic:2003, Nindos:Andrews:2004}. Observationally, calculation of the magnetic helicity is complicated since the vector potential ${\bf A}$ is not directly observable and is even not uniquely defined in terms of the directly-observable ${\bf B}$ field.  However, the current helicity, $H_{C}=\int_{V}{\bf B} \cdot (\nabla \times {\bf B}) ~dV$, is unique and its key components can be constructed from observations, and, therefore, $H_C$ has been extensively used in place of, or as a proxy for, $H_M$.

Detailed observations of the current helicity have once again exhibited a remarkable degree of temporal and structural coherency in the large-scale field \citep{Seehafer:1990, Pevtsov:Canfield:Metcalf:1995, Abramenko:Wang:Yurchishin:1997, Bao:Zhang:1998}.  These observations together have established the ``Solar Hemispheric Helicity Rule(s) (SHHR)". The SHHR primarily states that active regions in the northern hemisphere possess predominantly negative helicity, whereas active regions in the southern hemisphere have predominantly positive helicity.  This bias is cycle-independent but is not an absolute rule since it is obeyed by only $60-80\%$ active regions. Although the detailed temporal variation of adherence to the rule has not been established yet, various observational studies \citep{Bao:Ai:Zhang:2000, Bao:Pevtsov:Wang:Zhang:2000, Hagino:Sakurai:2005, Hao:Zhang:2011} at least seem to agree that the SHHR is violated more strongly at the transition between the cycles.  Modeling efforts that might contribute to the  explanation of the origin of the SHHR are the subject of this paper.

The observed structural properties of emerging magnetic flux at the solar surface has motivated the idea that  quasi-cylindrical bundles of relatively strong, toroidal magnetic flux, generally known as ``flux tubes",  rise from the solar interior due to magnetic buoyancy. Magnetic buoyancy can be thought of as the upwards force introduced by the presence of a magnetic field concentration in a stratified compressible fluid \citep{Parker:Buoyancy:1955}.  If the total pressure and temperature equilibrate quickly, as might generally be expected in this context, then the contribution of the magnetic pressure to the total pressure reduces the gas density locally where the magnetic concentration exists and produces the buoyancy force. Conceptually, the rise of magnetically-buoyant structures seems to fit the observations well.  For example, if the rise of a toroidal structure is not axisymmetric, then any upwards-arching of the rising magnetic structures (creating what is often called an  $\Omega$-loop) could eventually pierce the visible surface in such a way that the ``legs" of the loops then neatly explain the existence of sunspot pairs and their polarities.  Some interaction of the structure with the background global rotation during transit could further lead to the tilts of Joy's Law. However, despite these highly compelling conceptual ideas, the creation of such magnetic structures and the dynamics of their transport from the solar interior towards the solar surface is not adequately understood. 

The modeling of rising magnetic flux based solely on magnetic buoyancy (therefore generally ignoring convection and dynamo processes) can be broadly divided into two classes: (a) studies of the rise of preconceived magnetic ``flux tubes"; and, (b) studies of magnetic buoyancy instabilities. The first class of studies assumes the existence of a magnetic flux-tube-like structure without any dynamical inclusion of its originating process or field. It is important to note that, in this case, the ``flux tube'' is a purely arbitrary geometrical construction that is placed in a field-free  environment.  The environment may mimic certain solar-like conditions, such as the stratification, but the magnetic structure is totally isolated and disconnected from any larger-scale field (and the environment is often initially quiescent). Initial studies in this class \citep[e.g.][]{Moreno:1983, Moreno:1986, Choudhuri:1989, Dsilva:Choudhuri:1993, Fan:Fisher:Deluca:1993, Fan:Fisher:Mcclymont:1994, Caligari:Moreno:Schussler:1995, Longcope:Klapper:1997} focused on the ``thin flux tube approximation'' \citep{Spruit:1981} where a flux tube is purely a line with no cross-sectional area but has ascribed buoyancy, tension, and drag forces. Using appropriately-chosen parameters, these thin flux tube models can exhibit many of the characteristics of the solar observations, such as the correct latitudes of emergence and Joy's law, for example. Such agreements with solar observations have then been used to infer unobservable solar characteristics, such as the strength of the magnetic field at their region of formation  \citep{Choudhuri:Gilman:1987}.

Even though extremely useful in building initial intuition, these studies do not capture more complete dynamics, as was quickly discovered when finite cross-sectional flux tubes were modeled. For example, it was soon found via two-dimensional numerical simulations that, in order to maintain a coherent rise, finite-size flux tubes need to have a significant amount of magnetic field line twist in order to avoid being ripped apart by the trailing vortices generated in their wake \citep{Moreno:Emonet:1996}. This revealed the essential role of the twist, in this case, providing the necessary centrally-directed tension force required for a flux tube to remain cohesive.  The rise of finite-size flux tubes in these models is still driven by magnetic buoyancy, either from a pre-assigned non-equilibrium initial condition or via an instability \citep[e.g.][]{Schussler:1979, Schussler:Caligari:Ferrizmas:Moreno:1994, Moreno:Emonet:1996, Longcope:Fisher:Arendt:1996, Fan:Zweibel:Lantz:1998, Emonet:Moreno:1998}, although, as mentioned above, other dynamics, such as the wake dynamics, quickly play a significant role.  In three-dimensional numerical models of finite cross-sectional tubes, secondary instabilities were revealed that lead to kinking or arching structure \citep{Linton:Dahlburg:Fisher:Longcope:1998, Fan:Zweibel:Linton:Fisher:1998, Linton:Fisher:Dahlburg:Fan:1999, Fan:Zweibel:Linton:Fisher:1999, Fan:2001a, Fan:2001b}, very much reminiscent of the $\Omega$-loops envisaged as necessary to match the solar surface emergence observations.

The second general class of modeling studies has focused on the formation of magnetic flux structures from the instability of large-scale flux sheets.
The parcel argument for the concept of magnetic buoyancy briefly mentioned above can be rigorously formulated into an instability problem that generally reveals that instability can occur  when horizontal magnetic field increases sufficiently rapidly with depth (compared to the background entropy gradient)\citep{Parker:Buoyancy:1955, Acheson:1979}.  
Two- and three- dimensional simulations of horizontal magnetic flux sheets \citep[e.g.][]{Cattaneo:Hughes:1988,Matthews:Hughes:Proctor:1995,Hughes:Wissink:Matthews:1997,Wissink:Hughes:Matthews:2000,Vasil:Brummell:2008,Guerrero:Kaplya:2011} have exhibited that magnetic buoyancy instabilities evolve to form strong flux concentrations that are closely akin to the conceptual geometry of the isolated flux tubes described for the type (a) studies above.  The concentrations are pseudo-cylindrical, in the sense that they have a ``mushroom-like" cross-section perpendicular to the field lines that is substantially narrower than any variation along the field lines.  The long-wavelength variation down the initially-horizontal fieldlines corresponding to the most unstable modes naturally leads to  $\Omega$-like rising structures, as required by observations.  Only a few studies have included processes related to the origin and formation of the magnetic flux sheets that then subsequently give rise to these instabilities \citep[e.g.][]{Brummell:Cline:Cattaneo:2002, Cline:Brummell:Cattaneo:2003a, Cline:Brummell:Cattaneo:2003b, Cline:2003, Cattaneo:Brummell:Cline:2006, Kersale:Hughes:Tobias:2007, Vasil:Brummell:2008}. For example, in \citet{Vasil:Brummell:2008}, a forced, vertically-sheared, horizontal flow first creates a localized horizontal (``toroidal") magnetic sheet from seed initial vertical (``poloidal") field.  This sheet subsequently goes unstable to magnetic buoyancy instabilities \citep{Vasil:Brummell:2009} that again show the formation of strong magnetic flux-tube-like structures.

It should be noted at this point that there is a third class of studies, that of full global spherical convective dynamo models, where strong bands of toroidal magnetic field that potentially contain buoyant loop-like elements can be seen \citep{Brown:Browning:Brun:2010, Nelson:Brown:Brun:Miesch:Toomre:2011, Nelson:Brown:Brun:Miesch:Toomre:2013, Nelson:Miesch:2014}.  However, since the scale of these structures is very large, it is hard to relate these directly to ``flux tubes''.  Such structures could possibly be re-organized by near-surface processes into smaller-scale structures or could be the origin of the field for the other two types of investigations just mentioned. 

It is crucial to realize that, in the second class of magnetic buoyancy investigations (and especially the ones that include the layer formation process), the flux structures formed are not isolated magnetic entities, as is assumed in the first class. Instead, any magnetic structures formed are embedded in a large-scale background field.   Therefore such structures may be better referred to as magnetic flux concentrations. The rise of such  concentrations may be significantly different from the rise of isolated flux tubes in a field-free environment, as discussed in some detail in \citet{Cline:Brummell:Cattaneo:2003a}. For example, whether the enhanced connectivity of an embedded concentration in a large-scale background field helps or hinders rise is not understood.  Connectivity to the background field during the rise may create tension that opposes the rise.  On the other hand, connectivity could potentially also alleviate some of the issues with the rise of isolated structures, such as their need to conserve flux as the structure rises through a strong density stratification leading to ``ballooning" of the structure.  Interaction between the background field and the rising concentration could also certainly rearrange observable quantities of interest, such as the helicity of the structure.
This potential for different dynamics motivates the need for modeling that incorporates such possibilities when examining the causes of observable characteristics such as the SHHR.  We provide a first cut at this here in this paper. However, before we describe our efforts, we will outline other work that directly relates to the SHHR that has emerged from the previous categories of modeling.

Theories for the origin of the SHHR abound, but few are very complete dynamically or fit the observations in a highly compelling manner.  One might expect solar magnetic fields, in general, to have a handedness or chirality since much of our understanding of solar dynamo theory is based around the necessity of the existence of turbulent flows that are chiral in order to produce any mechanism that resembles a mean-field $\alpha$ effect  \citep{Steenbeck:Krause:Radler:1966}.  Of course, the Coriolis force is readily available to break symmetry to achieve this.  Many theories for the SHHR therefore rely on the existence of the Coriolis force either directly or indirectly.  Some authors have argued that the helicity in the SHHR is a direct result of the correlation of the fields in a mean field dynamo.  For example,  \cite{Gilman:Charbonneau:1999} explore the relationship between different mean-field dynamo models and the subsequently observed current helicity. 
Unfortunately, the results are very dependent on the exact formulation of the $\alpha$ effect and can produce mean fields that have the same correlations as the SHHR or the  opposite, depending on model choices.  Other models loosen the direct correlation with the $\alpha$ effect \citep{Gosain:Brandenburg:2019} somewhat by invoking scale-dependence, whereby the small-scales have the opposite sign of $\alpha$ to the large scales. 
Overall, however, such mean-field models can only provide the overall helicity of global axisymmetric fields and struggle to say something meaningful about anything that looks like an active region.  

Most other investigations therefore fall into the first category of magnetic buoyancy modeling (described as type (a) above), and have centered around how isolated, preconceived, rising flux tubes of some sort can `acquire' helicity during their transit to the surface\footnote{Note that most authors cited now refer to mechanisms for generating twist not helicity, leaving the direction of the toroidal field connecting the two to be understood from the underlying dynamo fields.  We refer to authors' results here in their language.}.   Rising isolated flux tubes writhe in response to the Coriolis force to create a tilt of the $\Omega$ loop in accordance with Joy's Law \citep{Wang:Sheeley:1991,Dsilva:Choudhuri:1993, Fan:Fisher:Mcclymont:1994}.  Since the tilt corresponds to the writhe of a flux tube, in order to conserve helicity, the tube acquires the opposite-signed twist.  Since the tilt was right-handed (left-handed) in the Northern (Southern) hemisphere, the twist is left-handed (right-handed) in accordance with the SHHR.  Unfortunately, Joy's Law tilts do not provide sufficiently strong enough twists to explain observed surface helicity values \citep{Longcope:Linton:Pervtsoc:Klapper:1999,Fan:Gong:2000}.  Furthermore, the latitudinal dependence expected in this case is not seen in observations \citep{Pevtsov:Canfield:1999,Holder:Canfield:Mcmullen:Nandy:Howard:Pevtsov:2004}. \citet{Wang:2013} attempted to circumvent this constraint by allowing the twist to be generated solely by the action of the Coriolis force on the expansion of legs of the $\Omega$ loop as it rises, concluding that sufficient helicity could be generated as long as the rise were slow enough.  \citet{Choudhuri:2003} postulated a different, highly conceptual model whereby the twist is acquired when an untwisted tube rises into a near-surface large-scale background poloidal field. In their ``mean-field circulation-dominated solar dynamo (CDSD)" model, this poloidal field originates from the decay of tilted active regions in a Babcock-Leighton type mechanism  \citep{Babcock:1961, Leighton:1969}.  The acquired twist is thus ultimately a result of the interaction of the dynamo field with rotation again. Inspired by the \citet{Choudhuri:2003} model, \citet{Chatterjee:Choudhuri:Petrovay:2006} and \citet{Hotta:Yokoyama:2012} have studied (via models and simulations respectively) the accretion of twist onto an initially untwisted flux tube in the presence of specific formulations of background field.

A general drawback of the above models is that they predict a fairly strictly-enforced SHHR and do not easily explain the observed weak (60-80\%) adherence to the rule and the significant inherent scatter therein (although \citet{Choudhuri:Chatterjee:Nandy:2004} does show some level of violation during the transition between cycles in the Choudhuri model, thanks to the correlation of the dynamo fields in the CDSD model).  One theory that does account for this, and is therefore perhaps the leading theory of the SHHR to date, is that of  \citet{Longcope:Fisher:Pevtsov:1998}, hereafter referred to as LFP. LFP invoke the action of rotationally-influenced convective turbulence on the transit of thin flux tubes through the convection zone.   In some sense, their ``$\Sigma$" effect is a small-scale version of the previous ideas, whereby the buffeting by Coriolis-influenced helical turbulence imparts net writhe to a thin flux tube that must be compensated by a net twist of opposite sign.  In this manner, the sign is again in accordance with the SHHR, and the turbulent nature leads to significant fluctuations and scatter, as required.  However,  this model has no obvious basis for temporal variations of the adherence to the rule over the solar cycle which becomes even more pronounced towards its end as found by various observations.

The biggest differences between our model for the SHHR (described below) and these previous models described above are that (i) we consider the dynamics of a finite-sized  concentration of magnetic field in a volume-filling field rather than an isolated magnetic entity, thereby more in accord with type (b) models and full dynamo simulations, and, (ii) our model allows for structures to be created with any initial helicity and then a selection mechanism  ``sorts" or ``filters" these to reveal the appropriate SHHR handedness rule in general but naturally then accompanied by a lot of scatter.    The SHHR in our case is derived from the initial configuration of the magnetic field and this filtering mechanism, and is not acquired during its transit.  Perhaps the closest previous model to our own is that of \citet{Choudhuri:2003} since it involves the interaction of a magnetic structure with an overlying dynamo-generated large-scale field.  However, the similarities stop there though since the expected locations of the two effects are entirely different, and, again, our model is a method of sorting helicities rather than creating (or acquiring) helicity.  It should be noted at this point that our mechanism can be readily combined with all the other mechanisms, and it is entirely possible, if not probable, that at least some of them play a significant role.  It is an interesting question as to what degree each contributes.

Preliminary results based on our model have been published in \citet{Manek:Brummell:Lee:2018} (hereafter referred to as Paper I:  see extended review in section \ref{sec:PreviousResults}). This preliminary work studied simply the effect of adding a background field of various strengths and directions to a canonical flux concentration.  However, these results clearly show that magnetic flux concentrations rising in the presence of even a relatively weak large-scale background field exhibit very different dynamics to the dynamics studied under the simplistic assumptions of isolated flux tubes rising in a field-free environment.  This work found that even weak background fields could quench the rise of the flux concentrations that would rise in the absence of background field.  More remarkably perhaps, the work discovered the selection mechanism mentioned above.  This selection mechanism, when applied to the solar context, agrees with many facets of the SHHR to a surprising degree of detail. In this followup paper here, we go beyond simply demonstrating the effect as in the original paper, and we focus on examining the robustness of these results to the many various assumptions of the model and the dependence on parameters. In particular, with the help of a detailed analytic model, we outline exactly when the bias that leads to the SHHR-like behaviour in this model will manifest in terms of the  relative strengths and configurations of the magnetic flux concentration and the background field. Further, we extend the work to create a synthetic SHHR map demonstrating the detailed agreement of this model to the observations. In particular, we specifically address the issue of significant scatter in the observations of SHHR and its temporal variation. 

\section{Model and Methods} \label{sec:ModelsMethods}

\begin{figure}
	\centering
	\includegraphics[width=6cm, height= 7cm]{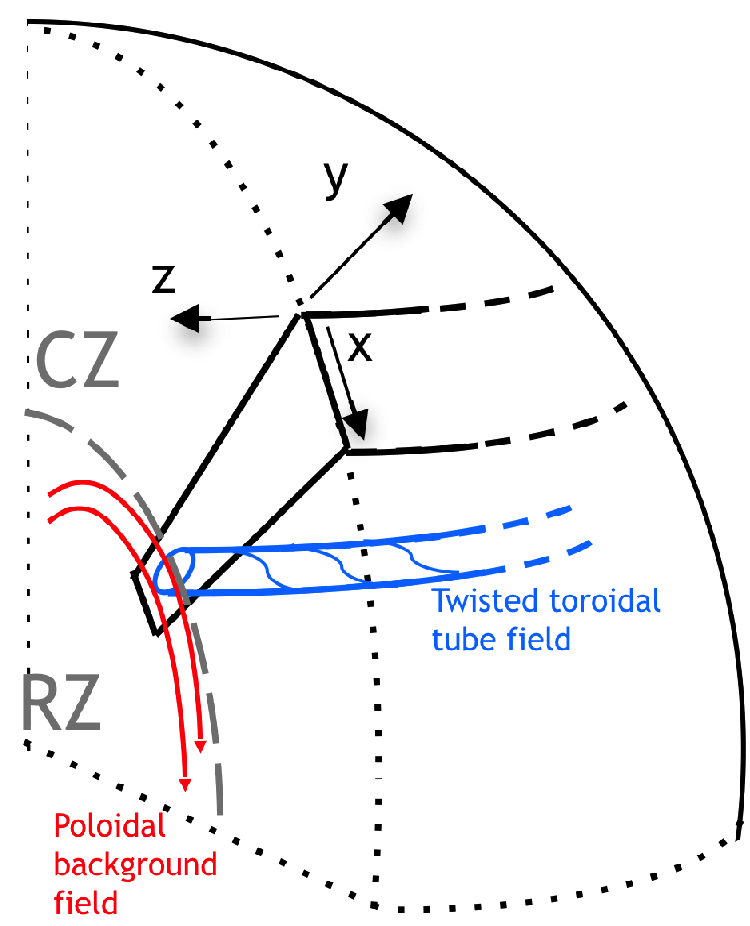}
    \caption{Cartoon sketch of the relation of the model to the Sun (adapted from Paper I).  Our 2.5 dimensional model relates the Cartesian directions $x$ to latitude, $y$ to height, and $z$ to longitude.  Vectors in the $z$ direction exist but the model is independent of this direction.  The domain spans the lower convection zone and the transition to the radiative zone. The model horizontal background field is related to deep, large-scale poloidal field.  The magnetic concentration is identified with twisted toroidal field structures, and is initially place deep in the stratification. }
    \label{fig:ModelSetup1}
\end{figure}

Our model essentially evolves a cylindrical flux structure (comprised of both axial and a locally-azimuthal magnetic field component so that the field is helical) embedded in a large-scale background magnetic field oriented horizontally and  perpendicularly to the tube axis. When comparing to the motivating solar application, the cylindrical flux structure should be thought of as toroidal (and therefore represents the typical idea of a twisted magnetic flux tube, soon after formation, in the deep interior), and the large-scale background field should be thought of as poloidal, representing the deep interior poloidal component of the dynamo field (see Fig.~\ref{fig:ModelSetup1}). The flux structure setup is very similar to many previous studies \citep[e.g.][hereafter referred to as HFJ]{Moreno:Emonet:1996, Hughes:Falle:Joarder:1998} but our model has the  crucial addition of the large-scale background field, making the flux structure a concentration rather than an isolated tube. We choose an  adiabatically stratified fluid layer covering 1.2 density scale heights so that it mimics roughly the region containing the upper tachocline and the lower $40\%$ of the convection zone. Even though no convection is present in our current simulation setup, we have chosen the initial background field profile to mimic what might be expected if indeed convection were to be present. That is, we concentrate the horizontal field near the bottom of the domain as if it had undergone magnetic pumping into the overshoot region at the top of the tachocline by the turbulent convection \cite[see e.g.][]{Tobias:Brummell:Toomre:2001}. Ultimately, however, our results turn out to be relatively insensitive to this choice of configuration. The more complicated problem with convection present will be addressed in later studies. We ignore the origins of these initial fields, assuming that they have already been formed by dynamo and instability processes, and study their evolution.  Note that these are not instability calculations, as the initial conditions are not in any equilibrium by choice; we imagine the start of our simulations to represent a portion of the later stages of a nonlinear instability simulation  \cite[e.g.][]{Vasil:Brummell:2008}.  Our aim is very simply to understand the effect of an volume-filling source background field on the rise of a pre-formed concentration under conditions akin to the deeper solar interior.  Other models \citep[e.g.][]{Chen:Rempel:Fan:2017} examine the emergence of field at the photosphere given initial conditions from the interior and it is one of our goals to build towards an understanding of the origin of the initial conditions needed for those simulations.

In order to the study the above-described model, we solve the equations of magnetohydrodynamics (MHD) using the publicly-available FLASH code \citep{Fryxell:Flash:2000, Dubey:FLASH:2013} in a Cartesian domain. The equations solved, in non-dimensional conservation form, are

\begin{subequations}
\label{eq:FLASH_Equations}
\begin{equation}
\dfrac{\partial \rho}{\partial t} + \nabla.(\rho \mathbf{v})=0,
\end{equation}

\begin{equation}
\dfrac{\partial \rho \mathbf{v}}{\partial t} + \nabla.(\rho \mathbf{vv}-\mathbf{BB})+\nabla p_{*}=\rho \mathbf{g} + \nabla.\tau,
\end{equation}

\begin{equation}
\dfrac{\partial \rho E}{\partial t} + \nabla.(\mathbf{v}(\rho E+p_{*})-\mathbf{B}(\mathbf{v.B}))=\rho \mathbf{g.v} + \nabla.(\mathbf{v}.\tau+\sigma \nabla T) + \nabla.(\mathbf{B} \times (\eta \nabla \times \mathbf{B})),
\end{equation}

\begin{equation}
\dfrac{\partial \mathbf{B}}{\partial t} + \nabla.(\mathbf{vB-Bv})=-\nabla \times (\eta \nabla \times \mathbf{B}),
\end{equation}

\text{where}

\begin{equation}
p_{*}=p+\dfrac{\mathbf{B}^{2}}{2},   
\end{equation}

\begin{equation}
E=\dfrac{1}{2}\mathbf{v}^{2} + \epsilon + \dfrac{1}{2}\dfrac{\mathbf{B}^{2}}{\rho}, 
\end{equation}

\begin{equation}
\tau = \mu \Big((\nabla \mathbf{v})+ (\nabla \mathbf{v})^{T} - \dfrac{2}{3}(\nabla.\mathbf{v} ){\bf {I}}\Big).
\end{equation}

\end{subequations}
\noindent
Here, $\rho$ is the density of a magnetized fluid, $\mathbf{v}=(v_x,v_y,v_z)$ is the fluid velocity, $\mathbf{B}=(B_x,B_y,B_z)$ is the magnetic field, $p$ is the fluid thermal pressure, $p_*$ is the total pressure (fluid and magnetic),  $g$ is the body force per unit mass, $\tau$ is the viscous stress tensor, $E$ is the specific total energy, $T$ is the temperature, $\sigma$ is the heat conductivity, $\eta$ is the magnetic resistivity,  $\epsilon$ is the specific internal energy, $\mu$ is the coefficient of viscosity (dynamic viscosity), and $\mathbf{I}$ is the unit (identity) tensor. Here, $\mu$, $\sigma$ and $\eta$ are chosen such that magnetic Prandtl number, $Pr_{m}=\dfrac{\sigma}{\eta}=1$ and Prandtl number, $Pr=\dfrac{\sigma}{\mu}=0.1$.

These MHD equations are solved in a local Cartesian 2D simulation box of   dimensions $ x \in [-1,1] $ and $ y \in [0,4] $, whose geometry relative to a spherical star is shown in Fig.~\ref{fig:ModelSetup1}. The simulations are 2.5-dimensional, in the sense that fields in the third ($z$) direction are included but that the dynamics are independent of this direction.  The non-dimensionalization can be thought of as specified by the relative size of the tube compared to the variation of the background hydrostatic state, which is a weak adiabatic polytropic stratification 

\begin{equation} \label{eq:PolytropicModel}
T = 1 + \theta y'; \hspace{0.1 cm} \rho = (1+\theta y')^{m};
\hspace{0.1 cm} p = (1+\theta y')^{m+1},
\end{equation}
where $ y'=4-y $.  Throughout this paper, we set non-dimensional temperature gradient  $ \theta = 0.25 $, and the polytropic index $m=1.5$ (the adiabatic value for a perfect gas with $\gamma = \dfrac{C_{p}}{C_{v}} = \dfrac{5}{3}$, where $C_p$ and $C_v$ are the specific heats at constant pressure and volume respectively). The polytropic hydrostatic model dictates that $\mathbf{g}=-g\mathbf{\hat{y}}=-\theta (m+1)\mathbf{\hat{y}}$. Note that our model is essentially at a fixed plasma $\mathbf{\beta}$. The variation of the the magnetic field configuration  does alter $\mathbf{\beta}$ somewhat, but for the parameters that we have surveyed, the variation is not substantial.

The magnetic initial conditions consist of a cylindrical concentration (which we still often  colloquially refer to as ``the tube") of radius $r \le R$ centered at $(x_c,y_c)$ (where $r$ is the local radius of the tube relative to its center: $r=\sqrt{(x-x_c)^{2}+(y-y_{c})^{2}}$), embedded within a background horizontal (but vertically-varying) field. Throughout this study, we choose to use $R=0.125$ and take the centre of the tube to be at $(x_c,y_c)=(0,0.5)$.  The axial ($z$ direction) field inside the concentration is constant and defines the scale for the amplitude of the magnetic field, so that $B_z=1$ there (although note that a Gaussian cross-section of axial field is used instead of this top hat profile to test robustness in Section ~\ref{sec:PreviousResults}).  To define the azimuthal field in the concentration, we use a potential function $ A_{z} $ to ensure the solenoidal condition is satisfied:
\begin{equation}
A_{z} = -q(x^{2} + (y-y_{c})^{2}) + K,
\end{equation}
where we choose $K$ such that the continuity of $ A_{z}=0 $ is ensured at the edge of the tube $r=R$.   The parameter $q$ defines the twist of the tube fieldlines and is a key parameter of the simulations. 
The magnetic field defining the concentration is therefore given by
\begin{equation} \label{eq:FluxTubeVector}
    \mathbf{B_{tube}} = 
    \left(-2q(y-y_{c}), ~2qx, ~1\right), \quad r \le R.
\end{equation}
The large-scale overlying background field in which the flux tube is embedded is aligned horizontally (in $x$) and we choose an exponential variation in $y$ to represent poloidal field perhaps more confined to the base of the convection or upper tachocline:
\begin{equation} 
    \mathbf{B_{back}} = (B_{back},0,0) =
    \left( 
    B_{s}~\text{exp} \Big(
    \dfrac{y_{c}-y}{2H_{B}} \Big)
    ,0,0  
    \right).
    \label{eq:BackgroundField}
\end{equation}

Here, $ B_{s} $ is therefore the strength of the background field relative to the initial magnetic field strength at the center of the flux tube, and $H_{b}$ is the scale-height of the exponentially-decreasing field. We vary both these parameters to explore the effect of the configuration of overlying field, from weak to strong, and from significantly confined to almost constant in the vertical. 
The total initial magnetic field in the concentration is therefore given by

\begin{equation}
    \label{eq:TopHatProfile_B}
    \mathbf{B_{in}} =\mathbf{B_{tube}} +  \mathbf{B_{back}} = 
    \left(-2q(y-y_{c}) + B_{s}~\text{exp} \Big(
    \dfrac{y_{c}-y}{2H_{B}} \Big), ~2qx, ~1\right), \quad r \le R,
\end{equation}
and outside the tube is simply $\mathbf{B_{out}}=\mathbf{B_{back}}$.
This setup with no background field ($B_{s}=0$) is similar to the case studied in HFJ with the parameter $\alpha=0$ (with the only differences being that we do not assume symmetry of the rising flux tube about the mid-plane, and we choose a slightly larger tube cross-section).

\begin{figure}
	\centering
	\includegraphics[width=6cm, height= 8cm]{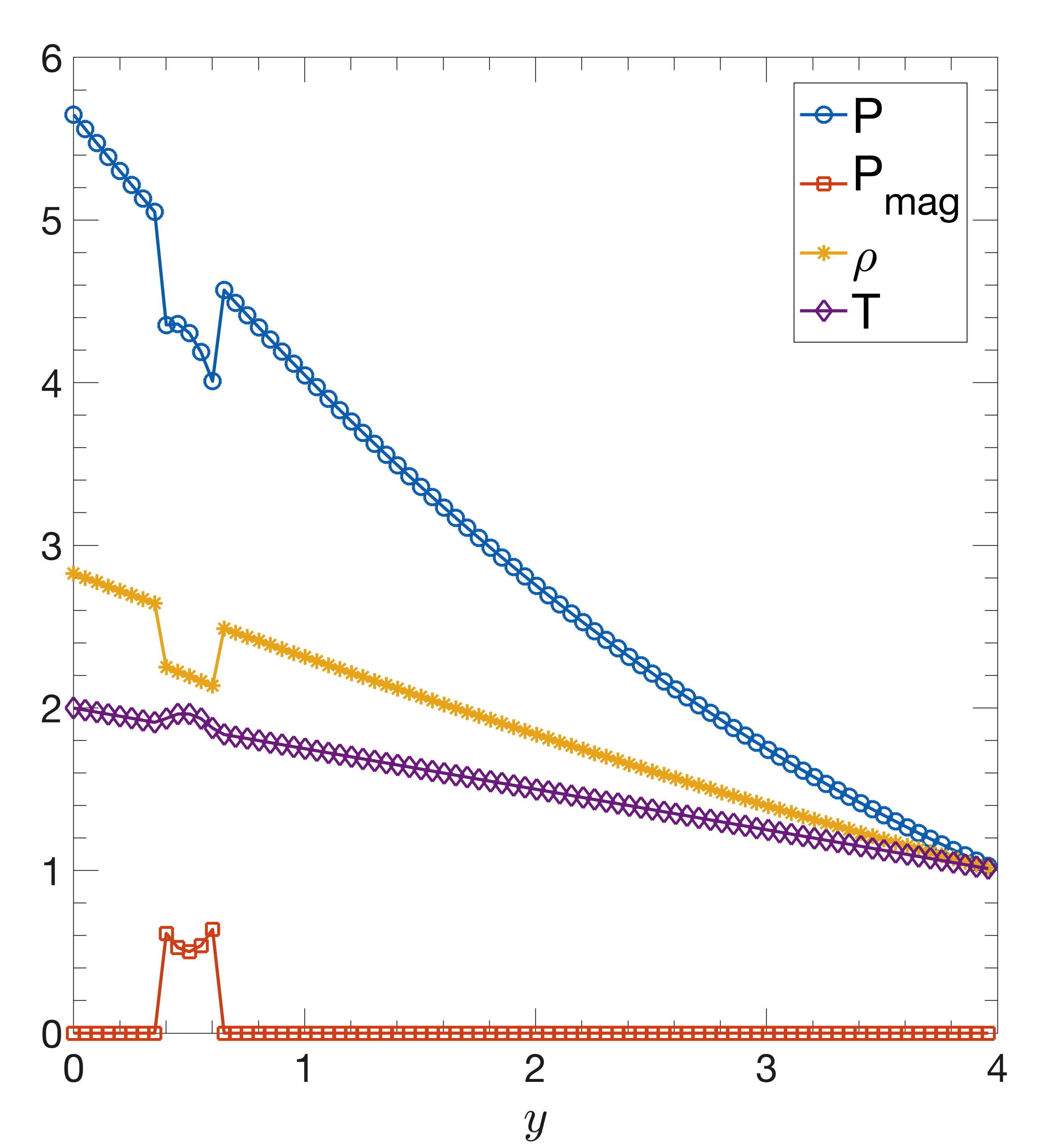}
    \caption{Vertical profiles of the initial thermodynamic conditions through the center of the magnetic flux concentration,  along $x=0$.}
    \label{fig:InitialThermo}
\end{figure}

On insertion of a magnetic field into a stratification, it is generally assumed that the total pressure equilibrates quickly making the total pressure continuous with the external conditions. The thermodynamics of the tube can then be specified such that there is a density anomaly with the tube less dense than the surroundings. However, there is no unique way of specifying these  initial conditions, as the effect of extra magnetic pressure can be accounted for by the  density, the temperature, or both combined. One formulation of the initial conditions that has been used by others (e.g. HFJ, \citet{Moreno:Emonet:1996}), and was used in Paper I, is to have the temperature continuous at the edge but varying inside the tube such that density and total pressure are merely a function of height (Fig.~\ref{fig:InitialThermo}).  We adopt this again here, but investigate the effects of varying the form of these initial conditions briefly as part of the robustness tests in this paper (see Appendix \ref{Appendix_InitialThermodynamicConditions}), finding that they have little bearing on the major results.

We choose the top and bottom boundaries of the simulation box to be fixed temperature, stress-free and impermeable to flow but not to magnetic fields, given respectively by\begin{equation}
T(y=0)=1 +\theta*4=2; ~~T(y=4)=1
\end{equation}
\begin{equation}
v_{y} = \frac{\partial v_{x}}{\partial y} = 0
\end{equation}
\begin{equation} \label{eq:BoundaryConditions}
 \frac{\partial B_{y}}{\partial y} = \frac{\partial B_{z}}{\partial y}  = 0 
\end{equation}

At the right and left boundaries of the simulation box, we implement Neumann boundary conditions
\begin{equation} 
\frac{\partial T}{\partial x} =\frac{\partial v_{x}}{\partial x} = \frac{\partial v_{y}}{\partial x}  = \frac{\partial B_{x}}{\partial x} =  \frac{\partial B_{z}}{\partial x} =  0 
\end{equation}
that allow outflow of the fields.  Whilst outflow magnetic boundary conditions are somewhat controversial \citep{Forbes:Priest:1987}, these conditions are commensurate with earlier studies of HFJ. In the current setup, the diameter of the tube is $1/8^{th}$ of the horizontal domain, and can rise vertically through 16 times its size thereby allowing significant room for unconstrained dynamics of a single tube. Hence, in this work, we only present dynamics far from the boundaries, and we have further checked that the results are robust to alternative choices of boundary conditions and other reasonable domain sizes.

\section{Previous Results: Dynamics Of a Flux Concentration} \label{sec:PreviousResults}

We briefly revisit the results of Paper I so that the key results, on which we are expanding here, are clear. However, so as not to be completely repetitive, and as one test of robustness, we change the radial profile of the initial magnetic flux concentration from the previous paper, keeping all other parameters the same. Paper I used a top-hat distribution for the axial field $B_{z}$, i.e. $B_{z} = 1$ for $r \le R$, where $R$ is the radius of the concentration.  Here, instead, we use a Gaussian profile in radius, where $B_{z}=B_{o} \exp(-r^{2}/R^{2})$ for $r \le 2R$, as also used in, for example, \cite{Cheung:Moreno:Schussler:2006}. We choose $B_{o}$ such that the net integrated $B_{z}$ inside the Gaussian flux tube is the same as the earlier studied top-hat distribution.

\begin{figure*}
	\includegraphics[width=\textwidth, height=21 cm]{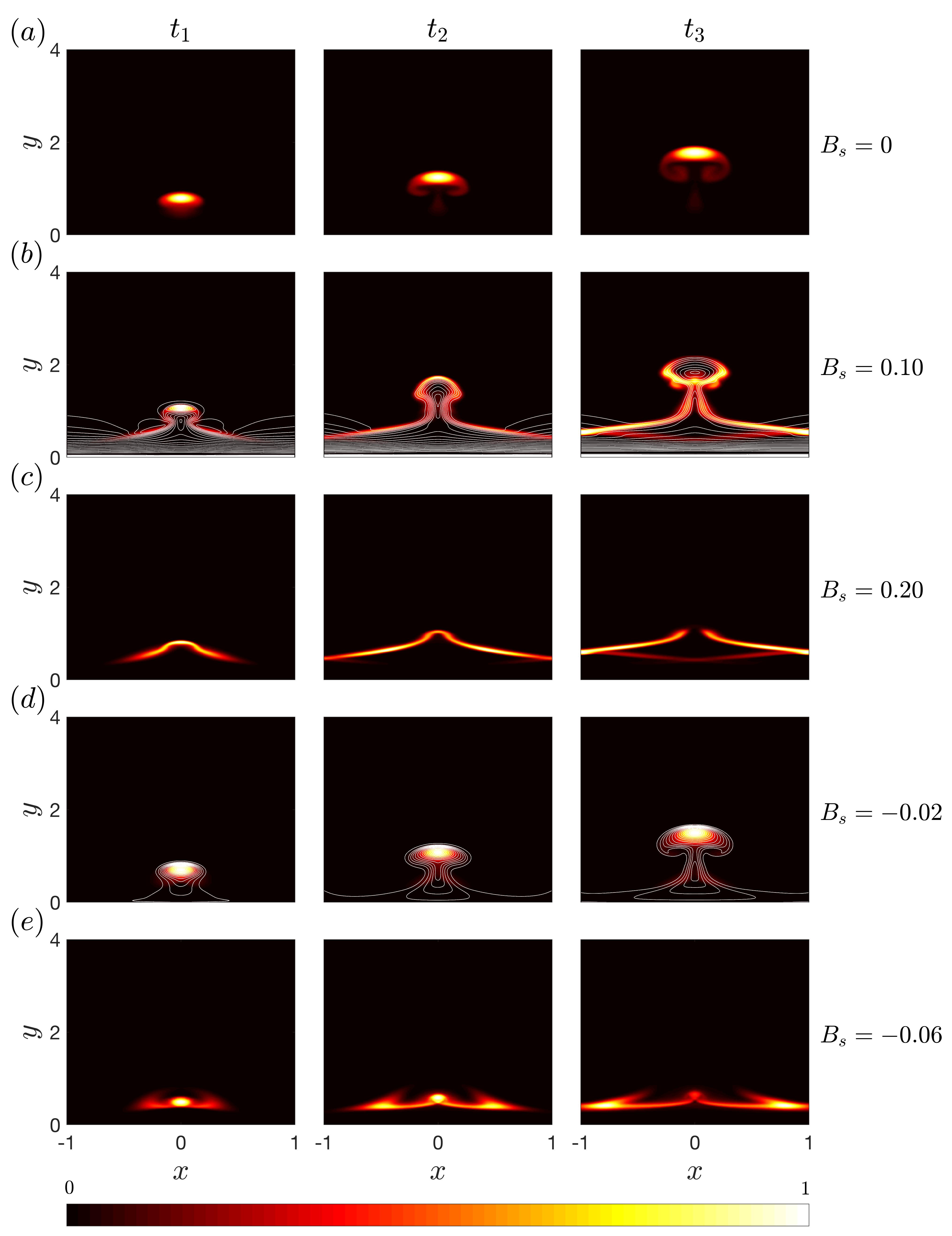}
    \caption{Intensity plots showing the evolution of (normalized) $B_{z}(x,y,t)$ for various background field strengths.  Shown in rows (a, b, c, d and e respectively) are  $B_{s}=0,~0.1,~0.2,~-0.02,~-0.06$.  For all cases,  $H_{B}=0.125$.  Three times are shown for each case.  In $(a,b,d,e)$ the times are $(t_{1},~t_{2},~t_{3}) = (5,~10,~15)$ but for (c) the times are $(t_{1},~t_{2},~t_{3}) = (2,~5,~8)$. Contours of $A_{z}$ have been added to panels (b) and (d). Panels (a, b and d) demonstrate cases of  successful rise, whereas in panels (c and e) the rise is suppressed. The initial axial magnetic field profile in each of these cases is given by Equation (\ref{eq:GaussianProfile}).}
    \label{fig:Figure2_Paper2}
\end{figure*}

Figure~\ref{fig:Figure2_Paper2} shows intensity plots of $B_{z}$ as a function of time for a number of simulations at varying background field strength and orientation, but otherwise fixed parameters (equivalent to Figure 2 of Paper I). The crucial results, which we now describe, are essentially the same, regardless of the form of the magnetic initial condition. Figure \ref{fig:Figure2_Paper2}a shows a canonical case exhibiting the characteristic rise of an isolated twisted magnetic flux tube in a field-free environment by the action of magnetic buoyancy. The flux tube rises coherently, forming a ``mushroom-like" structure, without any non-diffusive flux loss, a result found by many (e.g. HFJ). 

Paper I found that the introduction of background field significantly affects these rise dynamics, and this is also seen here in these new results in the subsequent panels of Fig.~\ref{fig:Figure2_Paper2}. In the presence of a background field of strength, $B_{s}=0.1$ (Fig.~\ref{fig:Figure2_Paper2}b), the rising flux structure experiences some initial flux loss due to transport down the overlaying background fieldlines, but rises in a qualitatively similar manner to the canonical case. Here, the rising flux structure is still reasonably coherent and is able to transport upwards a major portion of the initial flux in the flux tube. With a stronger background field strength, $B_{s}=0.2$ (Fig.~\ref{fig:Figure2_Paper2}c), the coherency of the initial flux structure is completely disrupted and no significant upwards flux transport is observed. A more fine-grained parameter study of varying background field strengths reveals that an overlying background field with a typical strength of $B_{s} \ge 0.16$ can completely halt the rise of an initially-buoyant twisted magnetic flux tube of the canonical type studied. The first surprising result of Paper I, and repeated here in a slightly different configuration, is therefore that a relatively weak background field (only 16\% of the initial tube strength) can totally suppress the buoyant rise of the structures.

The second, perhaps more surprising, result of Paper I, is that the quenching threshold depends on the relative orientation of the twist of the magnetic concentration (i.e. its local azimuthal field) and the background field. Paper I investigated this by keeping fixed twist, reversing the background field and varying the background field strength to find the value where rise was suppressed again, which we repeat here with our new Gaussian tube. Figure~\ref{fig:Figure2_Paper2}d shows the rise of a magnetic flux tube in the presence of an overlying large-scale background field of strength, $B_{s}=-0.02$ ($2\%$ of the initial axial field at tube center), oriented in the negative horizontal (x) direction, showing that this very weak strength does not affect the dynamics in any significant way. However, by slightly increasing the overlying field strength to $B_{s}=-0.06$ (Fig.~\ref{fig:Figure2_Paper2}e), the rise of the buoyant tube is completely suppressed and the coherent initial structure completely disintegrates. This surprising behavior sets another, even lower, background field strength threshold for the quenching of the flux tube rise with this field orientation.

Note that, in these particular investigations, the twist is always positive (i.e. the azimuthal field in the concentration is  oriented in a  counterclockwise fashion) and we then experimented by flipping the orientation of the background field. The symmetry of the problem allows us to make analogous conclusions for a case where the flux tube twist is flipped while keeping the orientation of the background field constant. That is, keeping the background field always positive, for example, then tubes with positive twist (where the azimuthal field forming the twist is aligned with background field at the bottom of the tube and counter to the background field at the top) are quenched when fields above $B_{s}=0.16$ are present, whereas tubes with negative twist (azimuthal field aligned with the background field at the top of the tube and counter to the background field at the bottom) are quenched at strengths above $B_{s}=0.06$. There are, therefore, certain intermediate background field strengths where one orientation of twist (relative to the background field) is suppressed whereas the other is not.  That is, there exists a ``selective rise regime (SRR)" in background field strength where a selection mechanism operates allowing structures of one twist to rise whilst suppressing the other. It so happens that this selection mechanism of tubes is remarkably commensurate with the SHHR in many respects, as was originally explained in Paper I and as we will outline again shortly. In this paper, and in particular in Section \ref{sec:AnalyticalResults}, we will investigate the mechanism, parametric dependencies and properties of this SRR in far greater detail.

\begin{figure}
	\includegraphics[width=\columnwidth]{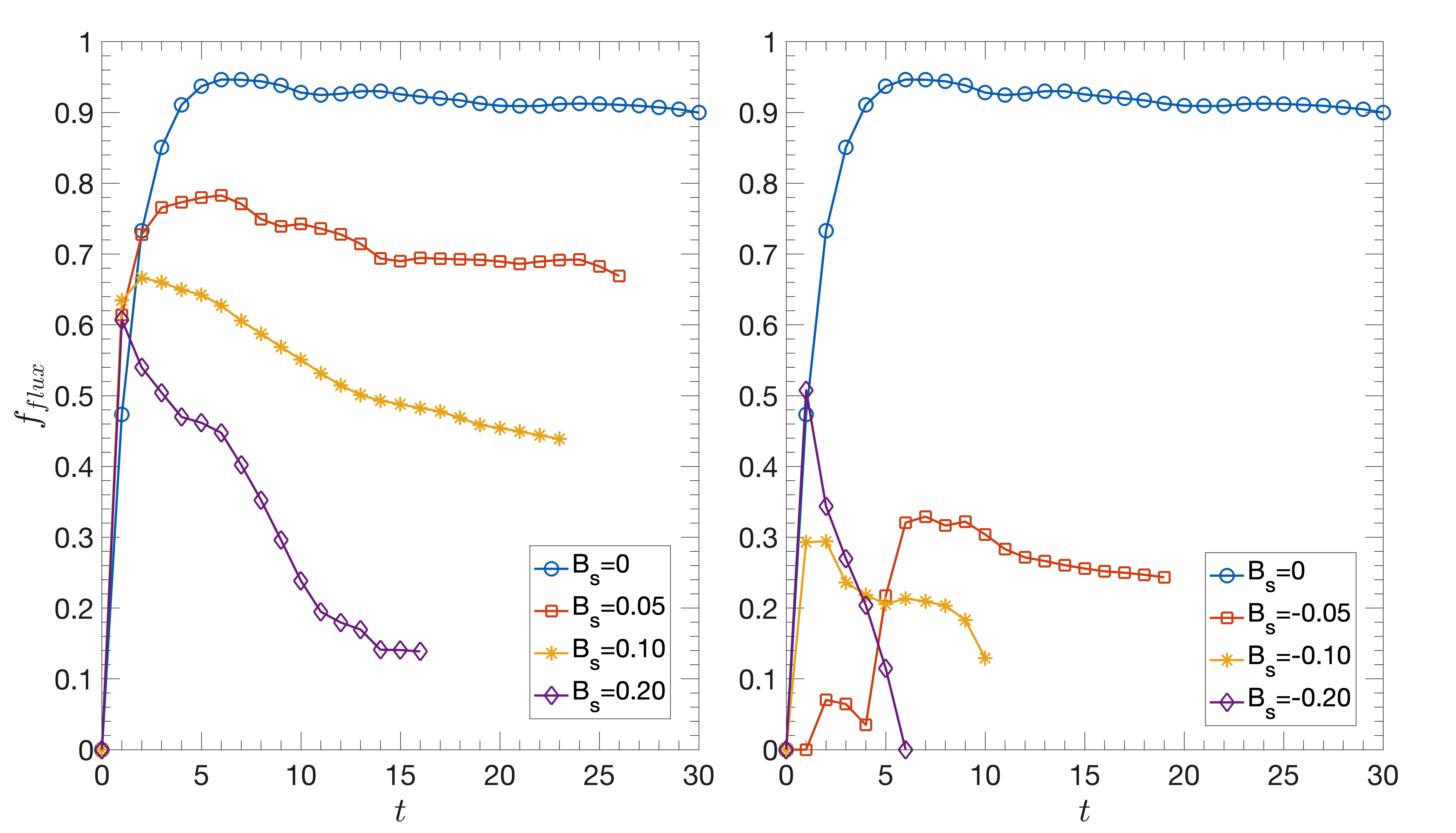}
    \caption{Rising flux fraction, $f_{flux}(t)$ for the initial Gaussian magnetic axial field profile. (a) $B_{s} \geq 0$, (b) $B_{s} \leq 0$ with $H_{B}=0.125$ and $v_{y,threshold}=0.02$. }
    \label{fig:MagneticICFluxFraction_figure}
\end{figure}

The qualitative picture of the rise and suppression of tubes in the presence of a background field as described by Fig.~\ref{fig:Figure2_Paper2} can be corroborated more quantitatively by calculation of the rising flux fraction, $f_{flux}(t)$ (as was done in Paper I for the top hat initial condition rather than the Gaussian one here). The rising flux fraction is given by

\begin{equation} \label{eq:flux_fraction_definition}
    f_{flux}(t) = \dfrac{\int \int B_{z,t}^{*}~dx~dy}{\int \int B_{z,0}~dx~dy}
\end{equation}

\noindent 
where $B_{z,0}=B_{z}(x,y,t=0)$ and $B_{z,t}^{*}=B_{z}({(x,y,t)~ \text{where}~ v_{y}(x,y,t) > v_{y,threshold}})$. Here, $v_{y,threshold}$ is a judiciously chosen velocity that tracks any flux rising upwards at or above this threshold velocity value.  A large flux fraction therefore indicates that a significant portion of the original axial flux in the flux tube is still rising, and lower flux fractions indicate that normal buoyant rise is being impeded. Figure \ref{fig:MagneticICFluxFraction_figure} shows the rising flux fraction as a function of time for various background field strengths and orientations with $H_{B}=0.125$. Panel $(a)$ of Fig.~\ref{fig:MagneticICFluxFraction_figure} corroborates the first surprising result that even a relatively weak background field strength ($\sim 20\%$) can suppress the rise of the flux tube. In this panel, the canonical result with $B_{s}=0$ can be seen to maintain the rise of almost all the flux (with only a slight diffusive loss) whereas $B_{s}=0.1$ reduces the rising fraction to less than $0.5$. Panel $(b)$ corroborates the second result regarding the effect of changing the orientation of the background field for a similarly twisted flux tube. Here, for a negatively-oriented background field (retaining a positive twist in the structure), the substantially more dramatic reduction of rising flux fraction by the same background field strengths is readily apparent. Note that both the panels show the rise only until the time that the flux tube either reaches the top of the simulation box or is completely stopped.

The reasons behind such observed dynamics were analyzed in Paper I.  We outline the basic ideas here, but these are investigated in much more detail in this paper in Section \ref{sec:Results} (including a mathematical model of the dynamics).   When background field is present, tension forces induced in the overlying background field as the tube attempts to lift it up during its rise, oppose that rise. Furthermore, the axial magnetic flux that provides the majority of the magnetic buoyancy for the rise can be transported (``drained") down the trailing background field, and the trailing vortices that normally drive the buoyant rise can be disrupted by the presence of the wrapped-around background field, causing further retardation. In the absence of background field, the net internal tension force in the flux tube is initially symmetric (independent of azimuthal angle, always pointing radially inward, therefore with no up or down component) and hence has no initial contribution in determining the rise characteristics, other than providing the coherency necessary for successful rise \citep{Moreno:Emonet:1996}. However, the introduction of a horizontal background field in the presence of a twisted flux concentration has the effect of adjusting the local $B_{x}$ at the leading and trailing edges of the flux tube differentially. The added background field can either enhance or detract from the contribution of the azimuthal field of the twist at either place, depending on the relative orientation of the two elements of the field.  For example, a positive background field enhances the azimuthal field at the bottom of a positively twisted tube and detracts from it at the top, whereas a negative background field acting on the same twist would enhance the top and decrease the bottom. Forces (internal to the flux tube) involving gradients of magnetic field, i.e., magnetic buoyancy and tension, are thus adjusted differently at the leading and trailing edges of the flux tube depending on the orientation of the background field. We find that an orientation of the background field and flux tube twist that enhances the local $B_{x}$ at the trailing edge of the flux tube and decreases it at the leading edge creates an asymmetric tension force in the tube that acts upwards, in concert with magnetic buoyancy forces. This enhancement acts against the retarding tension forces of the overlying field, and therefore a higher threshold of overlying field is required to suppress the rise. On the other hand, if the relative orientation of the background field and the twist has the effect of decreasing the local $B_{x}$ at the trailing edge of the flux tube and increasing it at the leading edge, then a net tension force is induced in the tube that acts downwards in concert with the retarding tension induced by rise in the overlying field. This increases opposition to the magnetic buoyant forces driving the rise, and therefore rise is more easily suppressed.

\begin{figure}
	\includegraphics[width=\columnwidth]{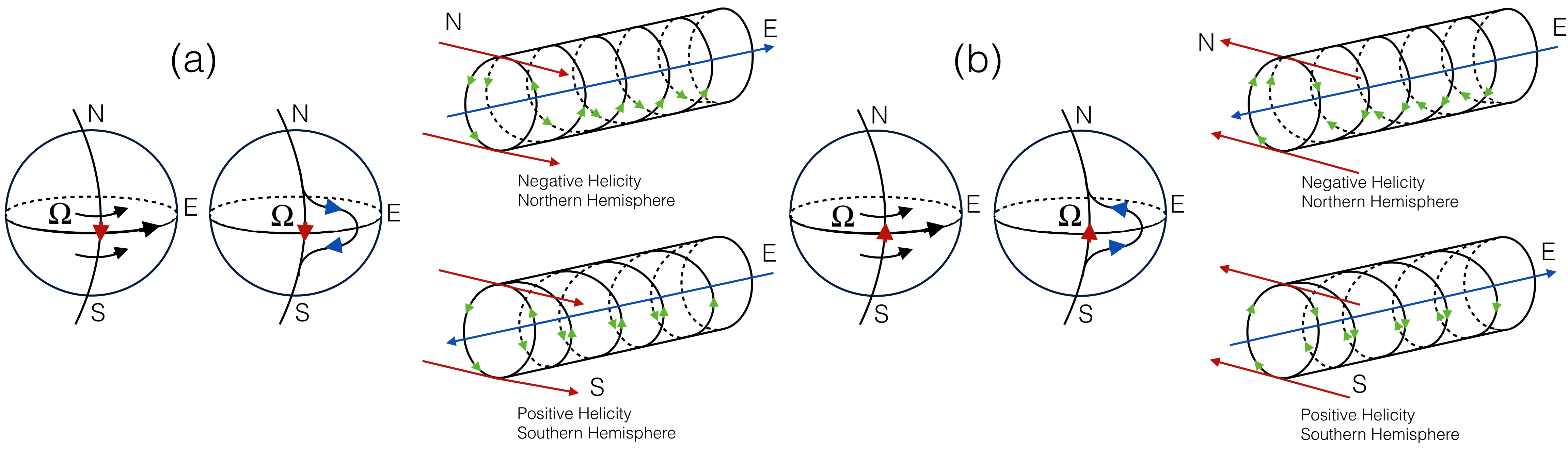}
   \caption{This figure shows how to translate our model results into the solar case.  Panels (a) and (b) show a pictorial representation of the first and second half of an arbitrary full 22-year solar cycle, respectively.
    The cartoon spheres depict the action of differential rotation on existing dynamo poloidal field (i.e. the mean-field $\Omega$ effect).  The dynamo poloidal field is represented with red arrow markers and toroidal field with blue markers. The differential rotation (exhibited as black arrow markers) stretches the poloidal field into toroidal field, leading to a certain orientation of the directions of the two field in each hemisphere (for each half of the cycle). The cylindrical cartoons show how the above oriented fields relate to the axial (blue, toroidal) and background field (red, poloidal) of our model. Furthermore, the tube cartoons show the twist (azimuthal, non-axial field, green markers) of structures that are predicted to rise  preferentially by our model. The resultant correlation of axial (blue) and twist (green) is commensurate with the SHHR. Figure adapted from Paper I.}
    \label{fig:3D_Cartoon_SHHR}
\end{figure}

Paper I showed that this selection mechanism, when applied to a solar context, had many qualities that agreed with the SHHR, but in particular the correct helicity parity. Figure~\ref{fig:3D_Cartoon_SHHR} shows how the model results translate to the solar scenario for a complete 22-year solar cycle where the fields reverse after 11 years. The azimuthal field direction (twist) of the flux concentration along with its axial field direction gives a certain sign of current helicity to the initial flux concentration. Note that our selection rule depends only on the relative orientation of the twist and the background field, and so the axial field direction, and therefore the current helicity, must be determined by the particular circumstances of the solar situation. Fig~\ref{fig:3D_Cartoon_SHHR}a shows the first half of a sample solar cycle defined by the N-S orientation of the large-scale poloidal field (red arrows). We assume, as is commonly the case in the current understanding of the generation of the solar large-scale fields, that deep in the solar interior, the action of differential rotation on the large-scale background poloidal field leads to the generation of strong toroidal flux sheets, which subsequently leads to more localised toroidal flux concentrations (likely via magnetic buoyancy instabilities). This assumption ensures that the orientations of toroidal axial field (blue arrows) of a flux concentration and the large-scale background poloidal field are correlated with each other. For example, for the first half of our sample cycle shown in Fig.~\ref{fig:3D_Cartoon_SHHR}a, the N-S orientation of the poloidal field leads to eastward toroidal field in the northern hemisphere and westward in the southern.  In the second half of our sample cycle (Fig.~\ref{fig:3D_Cartoon_SHHR}b), the poloidal field (red arrows) reverses (S-N) and the action of the differential rotation therefore produces westward toroidal field in the northern hemisphere and eastward in the southern. Based on the N-S orientation of the poloidal background field for the first half of the sample solar cycle in Fig.~\ref{fig:3D_Cartoon_SHHR}a, the selection mechanism of Paper I then requires that flux concentrations twisted in anticlockwise sense are more likely to rise (in either hemisphere) as they require a higher threshold of poloidal field strength to disrupt their dynamics. This anticlockwise twist then paired with the different axial toroidal field direction in each hemisphere then dictates which helicity has a preferential rise.  A positive correlation between the twist and axial toroidal field direction would lead to a positive helicity flux concentration and similarly a negative correlation would lead to a negative helicity flux concentration. It can be seen in Fig.~\ref{fig:3D_Cartoon_SHHR}a that flux structures with negative helicity in the northern hemisphere and positive helicity in the southern hemisphere are the ones that are more likely to rise. For simplicity of understanding, Figure~\ref{fig:3D_Cartoon_SHHR} shows these cases -- the ones that require a higher threshold of background field to halt/breakup the rise of a flux concentration, i.e., the configurations that are preferred in the sense that they are more likely to rise. It can be seen that Figure~\ref{fig:3D_Cartoon_SHHR}a is in agreement with the ``Solar Hemispheric Helicity Rule". Furthermore, in the second half of the cycle, shown in Figure~\ref{fig:3D_Cartoon_SHHR}b, both the poloidal field and the toroidal field flip direction (sign) in each hemisphere, thereby preserving the helicity of the structure that has preferential rise. This selection mechanism is therefore commensurate with the parity rules of the SHHR and its invariance over the full solar cycle.

The mechanism discovered also points to a potential explanation for the large scatter found in the SHHR observations, that was not examined in Paper I but is explored in detail in this paper. The selection only happens for a range of the relative strengths of the twist of the structure and background field strengths (which we call the SRR).  We might expect either of these contributions to vary outside of this range thereby allowing violations to the rule. 
We investigate, and also synthetically reproduce, this scatter in more detail later in this paper.

The above results originally reported in Paper I (and here repeated for a different axial field profile of the tube-like structure) were obtained from a small number of experiments based around a canonical setup. As the main purpose of this paper, we now proceed to quantify the robustness and parametric dependence of the SRR in detail through a much broader range of simulations (Section \ref{sec:Results}).  We also, in particular, expand the understanding of the mechanism by producing an analytical model (Section \ref{sec:ForceAnalysis}). Furthermore, to create a more realistic comparison between the model ideas and the observations of the SHHR, we then extend the single flux structure simulations to Monte Carlo (MC) simulations containing multiple tubes with randomly-generated properties, in order to examine the helicity distribution of the emerging flux concentrations (Section \ref{subsec:MultipleMonteCarlo}).

\section{New results:  Parametric dependence of the Selective Rise Regime (SRR)} \label{sec:Results}

\subsection{A broader survey of $q-B_{s}$ space at fixed $H_B$} \label{sec:AnalyticalResults}

We define our SRR as the region in parameter space where one relative orientation of twist and background field rises successfully but the reverse orientation does not.  Seeing as the dynamics of the selection mechanism depend on the interaction between the background field and the twist of the tube, it might be expected that the SRR is most strongly dependent on $q$ and $B_{s}$, and we concentrate on this first, fixing for now the other main parameter, the scale height of the background field configuration, at $H_{B}=0.125$ (and, as always, keeping the other parameters at their canonical values: $m=1.5,~R=0.125,~\theta=0.25$).  It is perhaps easiest to think of the SRR as the parameter regime where one sign of twist rises but the other does not at fixed background field strength and orientation.  Note, however, that our original simulations fixed the twist and examined the effect of reversing the field. 
The SRR is, therefore, really a two-dimensional region of $q-B_s$ parameter space (for fixed $H_B$) and we investigate this here.

\begin{figure}
    \centering
    \includegraphics[width=11cm, height=8cm]{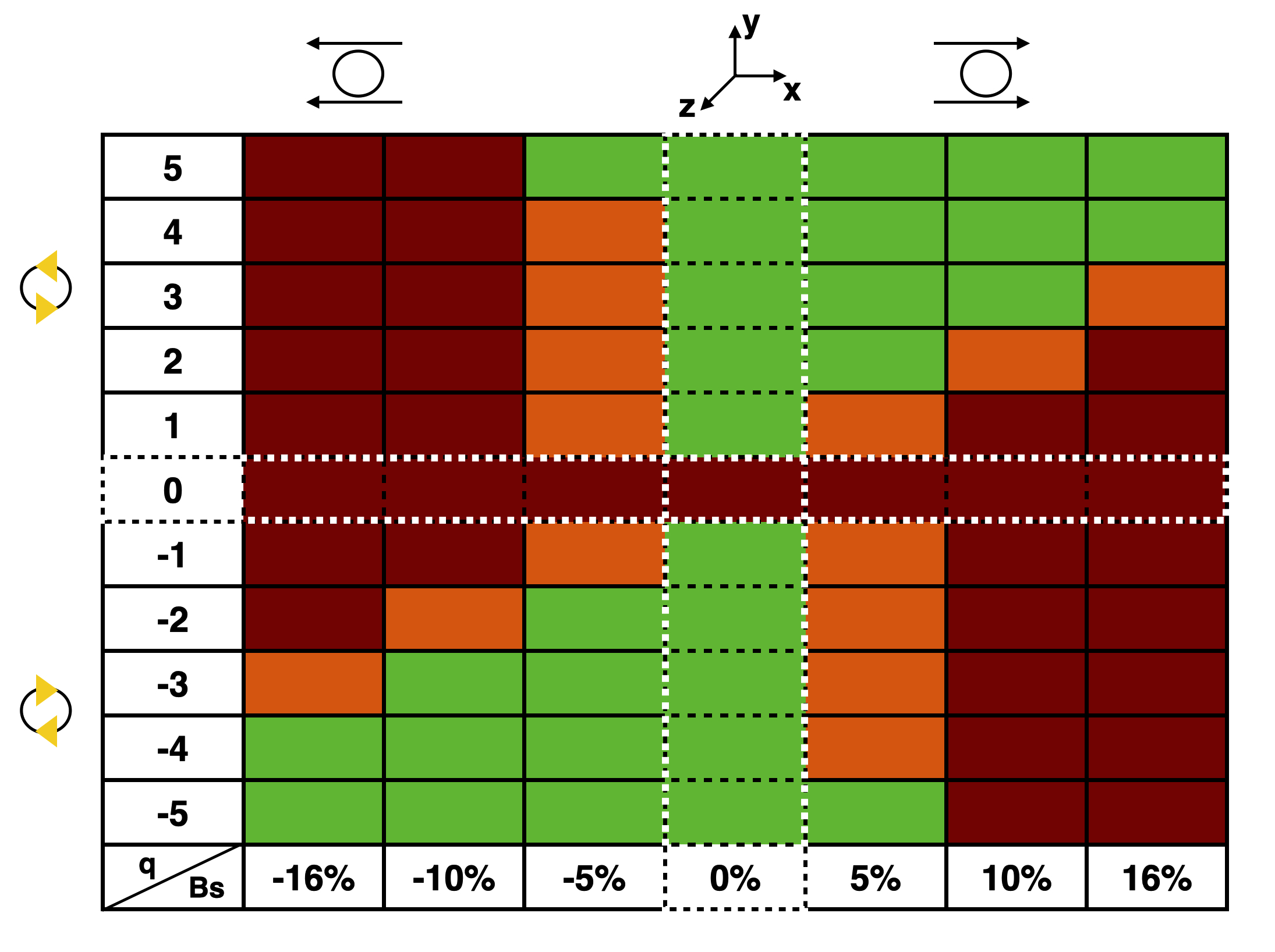}
    \caption{A map of rise success/failure in $q-B_{s}$ space for fixed $H_B=0.125$. Background field strengths $B_s$ are given as percentages of the axial field strength of the magnetic flux structure. Background field and flux tube twist orientation by sign are shown schematically for clarity. Red (darkest) color signifies a definite failed rise of the flux structure, whereas the green (lightest) color signifies a definite successful rise. Orange (intermediate shade) signifies an intermediate result, where the structure may rise but the dynamics are noticeably different than other rising cases. Simulations lying inside the white dotted lines are either cases with negligible background field $B_s \sim 0$ or cases where the twist is so low $q \sim 0$ that structures would not rise coherently. 
    }
    \label{fig:q_Bs_SurveyChart}
\end{figure}

Figure \ref{fig:q_Bs_SurveyChart} shows a much broader survey of results in the $q-B_s$ parameter space than was shown in Paper I. The original work of Paper I would correspond to a single row at $q = 2.5$ (not actually shown in this new figure). Schematic indications of the crucial twist and background field orientations are shown in the border of the table, along with their values, for ease of interpretation of the signs. The figure shows a matrix colored at each $q-B_s$ value according to whether a simulation at those values shows that the flux concentration clearly rises (green), clearly fails to rise (red), or something less easily determined (orange).  The determination in each simulation is done using visualisations and the flux fractions, as exemplified in Figures  \ref{fig:Figure2_Paper2} and \ref{fig:MagneticICFluxFraction_figure}. The relatively complicated intermediate dynamics represented by the orange colored cases are examined in more detail in Section  \ref{subsec:MultipleMonteCarlo}. Swapping both the sign of $q$ and $B_s$ results in the same dynamics (equivalent to just viewing the same system from the other end of the flux tube), and so the table is diagonally-symmetric across the origin. All four quadrants of the matrix have been included for ease of using the table, but this symmetry emphasizes that the selection mechanism only depends on the {\it relative} direction and strength of $B_s$ and the twist, $q$.  We, therefore, explain one half of the survey chart ($Bs > 0$) since analogous interpretations can be drawn for the other half. Note that in our model setup, when $q$ changes sign, the current helicity of the tube changes sign, since $B_z$ remains fixed in the positive $z$ direction.  This is irrelevant to the selection mechanism, since it depends solely on twist, not helicity. For relevance to the solar case, however, the selection mechanism can be cast in terms of the helicity, since, there, the sign of the background field and the tubes axial field are correlated, according to our understanding of the dynamics of the solar dynamo. The cases within the white dotted lines are special cases. For $q \lesssim 0.25$, all cases fail to rise coherently regardless of the  background field strength because some initial twist is required to maintain the coherency of the tube \citep{Moreno:Emonet:1996}.  For all other $q \gtrsim 0.25$, but zero background field, the tube rises successfully.

For any fixed $B_{s}$, the table in Figure \ref{fig:q_Bs_SurveyChart} clearly shows that an asymmetry exists between positive and negative $q$. For $B_s>0$, the transition from red (unsuccessful rise) to green (successful rise) for negative $q$ occurs (if it occurs at all in the cases surveyed) at far larger $|q|$ than for positive $q$.  By symmetry, for $B_s<0$, the roles are reversed with the $q>0$ boundary occurring at much larger $|q|$.  The different $|q|$ at which the red-green (unsuccessful-successful) transition occurs (say $|q_-|$ for the negative $q$ and $|q_+|$ for the positive $q$) delineate the SRR at that $B_{s}$. 
The SRR can therefore be defined from the simulation data at any $B_s$ as the set 
\begin{equation} \label{eq:GoldilocksZone_data}
    \text{SRR} = \{q ~~{\text s.t.} ~~|q| \in [\text{min}(|q_{-}|,|q_{+}|),~\text{max}(|q_{-}|,|q_{+}|)]\}
\end{equation}
Within this set, $q$ of one sign rises and the other does not. 
It can easily be verified from Figure
\ref{fig:q_Bs_SurveyChart} that, for $B_{s}>0$, it is structures with $q>0$ that rise if their $q$ lies in the SRR found at that $B_s$, whereas for $B_{s}<0$, it is the structures with $q<0$ that are the ones that rise. 

Figure \ref{fig:q_Bs_SurveyChart} shows that the boundaries of the SRR, given by $|q_-|$ and $|q_+|$, vary with $B_s$.  For higher $|B_s|$, stronger $q$ is clearly required to overcome the background field in order to rise, and the SRR becomes defined by higher $|q|$ values and appears to widen in extent of $|q|$.  These issues are investigated in detail via a numerical model in the next section.  

At this point, it is worth noting that there are some physical limits on what we might expect for reasonable values of $|q|$, despite the fact that we have no direct observational evidence from the solar interior to help us out in this matter. Firstly, in our model, structures with $|q| \gtrsim 12$ are unphysical, since then the azimuthal field is sufficiently strong  that the associated magnetic pressure induces a negative density in the initial conditions for the structure \footnote{This limit is actually not constant but instead a very weakly decreasing function of $B_s$, and is shown later as the dashed line in Fig. \ref{fig:Hb125_SRR_Explained}}.  Secondly, structures with $|q| \gtrsim 8$ are likely unstable to kink instabilities (according to \citet{Linton:Longcope:Fisher:1996}, whose three-dimensional model has a different field configuration from our two-dimensional model, but is likely a reasonable guide).  These issues informed our choice of range in $q$ in these simulations.  Similarly, we have no direct observational data for the strength of solar interior magnetic fields, but it seems reasonable to presume that the background poloidal field strength is only a fraction of the strength of any rising flux structure.

Ultimately, the $q-B_{s}$ diagram in Figure \ref{fig:q_Bs_SurveyChart} exhibits compactly the results of a selection mechanism and therefore illustrates clearly what we have referred to as the SRR, the parameter regime in which the selection occurs.  The overall conclusions from the simulation data presented in 
Figure \ref{fig:q_Bs_SurveyChart} may be that an SRR of some extent occurs for almost all reasonably physical parameter values.  This seems to imply the existence of an overall bias for the rise of flux structures, even if they have significant variance in their magnetic composition.  In general, a prevalence of positively twisted structures might be expected to emerge from a distribution of structures in the presence of a positive background field, and a prevalence of negatively twisted structures might be expected in a negative background field.  We verify these ideas in Section \ref{subsec:MultipleMonteCarlo}, but first we turn to an analytical model that defines the SRR explicitly in terms of the parameters, allowing a more fine-grained exploration and understanding of the parameter space.

\subsection{An analytical model of the dynamics}  \label{sec:ForceAnalysis}

Our previous paper, Paper I, provided some brief insight into the reasons for the varying dynamics associated with different orientations of the fields and tubes.  Here, we extend those ideas into a detailed model that can explain and predict the $q-B_{s}$ behaviour discussed above, and the dependence on other parameters. To formulate the model, we carry out a theoretical analysis of the key magnetic forces acting on the flux concentration in the vertical $(y)$ direction as it initially begins to rise. Our model is based on the following ideas. Firstly, we have in mind that the rise of a tube-like concentration through the overlying background field is generally hindered by tension induced in the background field as it is wrapped around the structure and gets lifted and stretched upwards.  Secondly, originating from the ideas in Paper I, we expect that the introduction of the background field induces differential values of the internal tension and buoyancy forces across the tube-like concentration which can lead to net forces that can help or hinder the rise. To create a model that predicts whether rise is successful or not, we calculate the differential buoyancy and tension forces from the initial conditions at $t=0$ and then estimate the impeding tension forces from the overlying field that are induced as the structure rises.  A balance of these forces should delimit the bounding case between successful and unsuccessful rise.  Note that, since the selection mechanism and all the results discussed  so far are robust to different formulations the magnetic initial conditions (see Section \ref{sec:PreviousResults} and Appendix \ref{Appendix_InitialThermodynamicConditions}), in order to make analytic integration of the net forces easier, we revert back to the ``top-hat" magnetic initial conditions ($B_z=1$ for $r \le R$ in the concentration; see Eqn \ref{eq:TopHatProfile_B}) as used in Paper I rather than the Gaussian cross-section used in Section \ref{sec:PreviousResults}.

The buoyancy force in the y-direction, $F_{buoyancy,y}$, including its magnetic contribution, is generally thought of as the driving force for the rise of the concentration. We have deliberately set up our problem so that the presence of the magnetic field of the concentration  produces a density (and temperature) perturbation (under the assumption of fast equilibrium of the total pressure; see Appendix \ref{Appendix_InitialThermodynamicConditions}, Case 2) that will drive the rise.  The addition of a background field that varies with height can create a perturbation to the vertical buoyancy force that is asymmetric about the center of the tube and therefore has a net contribution. Ultimately, however, we will find in the following calculation that the magnetic contribution to the buoyancy force is dominated by  the axial field of the tube, and that this differential perturbation is unimportant (for the buoyancy force, although it will be crucial for the tension force).  The net buoyancy force in the vertical (y) direction over the circular region of the tube-like concentration in our Cartesian coordinate system is given by 

\begin{equation} \label{eq:Buoyancy_Force_Definition}
    F_{buoyancy,y}=  \int_{-R}^{+R}  \Bigg[ \int_{-\sqrt{R^{2}-y'^{2}}}^{\sqrt{R^{2}-y'^{2}}} \Bigg(-\dfrac{\partial}{\partial y'} \Big(p_{gas,in}+\dfrac{B_{eff}^{2}}{2} \Big) + \rho_{gas,in} g\Bigg) dx \Bigg]  dy'. 
\end{equation}

\noindent
where $\mathbf{g}=\theta (m+1)\mathbf{\hat{y}}$ from equation (\ref{eq:PolytropicModel}) and $y'=y-y_{c}$.  This latter translation does not affect the analytical calculations mentioned below and hence for clarity, we drop the dashes in the subsequent calculations from $y'$ and simply use $y$.  Here, $B_{eff}$ is the total magnetic field inside the flux tube accounting for both the field of the magnetic flux structure and the background field; $p_{gas,in}$ and $\rho_{gas,in}$ are the gas pressure and density inside the tube.
Assuming total pressure balance between the inside of the tube and the outside, we have
\begin{equation} \label{eq:PressureBalance}
    p_{gas,in} + \dfrac{B_{eff}^{2}}{2} = p_{gas,out} + \dfrac{B_{back}^{2}}{2},
\end{equation} 
where $p_{gas,out}$ and the corresponding $\rho_{gas,out}$ are the gas pressure and density outside the flux tube in the stratified background given by the polytropic model (Equation \ref{eq:PolytropicModel}). The density inside the flux tube, $\rho_{gas,in}$, is then given in terms of $\rho_{gas,out}$ adjusted by the magnetic field (as described in more detail by the initial conditions outlined in Appendix \ref{Appendix_InitialThermodynamicConditions}, Case 2):

\begin{equation} \label{eq:FluxTubeInitialDensityProfile}
    \rho_{gas,in}(y)=\rho_{gas,out}(y)-\dfrac{1+4q^{2}R^{2}-4qyB_{back}}{2(1+\theta (4-(y+y_{c})))}.
\end{equation}

\noindent 
The factor in the denominator related to the initial polytropic temperature profile above, $T_{gas,out}=1+\theta (4-(y+y_c))$, varies only by a small fraction across the vertical extent of the flux tube and can be assumed to be roughly constant, equal to the average value, $T_{avg}$.  This makes the analytical calculation of the integrals in Equation (\ref{eq:PressureBalance}) more tractable. Plugging (\ref{eq:PressureBalance}) and (\ref{eq:FluxTubeInitialDensityProfile}) into (\ref{eq:Buoyancy_Force_Definition}) and using the polytropic nature of the outside gas, the net buoyancy force equation reduces to

\begin{equation} \label{eq:NetBuoyancyForceExpanded}
    F_{buoyancy,y} = \int_{-R}^{+R}  \Bigg[ \int_{-\sqrt{R^{2}-y^{2}}}^{\sqrt{R^{2}-y^{2}}} \Bigg(  -\dfrac{\partial}{\partial y} \Big(\dfrac{B_{back}^{2}}{2} \Big) -\theta(m+1) \Big(\dfrac{1+4q^{2}R^{2}-4qyB_{back}}{2T_{avg}} \Big)  
    \Bigg) dx \Bigg] dy.
\end{equation}
The final term in the integral is related to the cross term $(\sim B_{x}B_{back})$ and yields a difficult term in the integration.  We drop this term here, expecting it to be relatively insignificant, and indeed verify that this is the case later. 
In order to integrate the first term in equation (\ref{eq:NetBuoyancyForceExpanded}), we use Equation (\ref{eq:BackgroundField}) to substitute

\begin{equation} \label{eq:BackgroundFieldSquare_Derivative}
    \dfrac{\partial}{\partial y} \Big( \dfrac{B_{back}^{2}}{2} \Big) = -\dfrac{H_{B}}{2} B_{back}^{2} = -\dfrac{H_{B}}{2}{B_s^2} \exp\left(\dfrac{-y}{H_B}\right),
\end{equation}
and then, after integrating equation (\ref{eq:NetBuoyancyForceExpanded}) once with respect to $x$, we further use the definition of modified Bessel function of first order, 

\begin{equation} \label{eq:ModifiedBesselFunction_FirstOrder}
    I_{1}(z) = \dfrac{z}{\pi} \int_{-1}^{1} (1-t^{2})^{1/2} e^{-zt} dt,
\end{equation}
with $t=y/R$ and $z=R/H_{B}$. The result of the overall integration of Equation (\ref{eq:NetBuoyancyForceExpanded}) (excluding the cross term) is then

\begin{equation} \label{eq:NetBuoyancyForce_QuadraticQ}
    F_{buoyancy,y} = \pi B_{s}^{2}H_{B}^{2}R I_{1}(2\tau) + \dfrac{\theta (m+1) \pi R^{2}}{2T_{avg}} (4q^{2}R^{2}+1),
\end{equation}
where $\tau = R/(2H_{B})$.  Equation (\ref{eq:NetBuoyancyForce_QuadraticQ}) shows that the net buoyancy force has components that depend on the strength of the background field (through the first term and $B_s^2$) and the tube field strength (through the second term, dependent on $q^2$ and a constant value for the axial field).  When evaluating this expression for our parameters later, we will see that the second term clearly dominates.

When a background field is included, the role of the internal tension force in the overall dynamics can be equally as important as the buoyancy force in determining the net upwards driving force. When there is no background field ($B_{s}=0$), the internal tension force in the tube is symmetric and acts towards the center of the tube. Hence, there is no net force in any direction at $t=0$ and tension serves only to maintain the coherency of the tube in this case.  On the introduction of a background field that varies in the vertical, this internal tension force becomes asymmetrical around the horizontal mid-line of the tube, and, as pointed out in Paper I, the net vertical force induced can be upwards, thereby aiding the rise, or downwards, thereby hindering the rise. We denote this internal tension force in the vertical (y) direction by $F_{tension,y}$, given by 
\begin{equation} \label{eq:TensionForce_Definition}
    F_{tension,y} = \int_{-R}^{+R} \Bigg[ \int_{-\sqrt{R^{2}-y^{2}}}^{\sqrt{R^{2}-y^{2}}} (\mathbf{B\cdot\nabla})B_{y} ~dx \Bigg] dy = \int_{-R}^{+R}  \Bigg[ \int_{-\sqrt{R^{2}-y^{2}}}^{\sqrt{R^{2}-y^{2}}}  \Big(B_{x} \dfrac{\partial B_{y}}{\partial x}+ B_{y}\dfrac{\partial B_{y}}{\partial y}  \Big)dx \Bigg] dy,
\end{equation}
where $y$ has been translated as before. With the chosen magnetic initial conditions given in equation (\ref{eq:TopHatProfile_B}), $\dfrac{\partial B_{y}}{\partial y} = 0$ and so, at $t=0$,

\begin{equation}
    F_{tension,y} = \int_{-R}^{+R}  \Bigg[ \int_{-\sqrt{R^{2}-y^{2}}}^{\sqrt{R^{2}-y^{2}}}  B_{x}\dfrac{\partial B_{y}}{\partial x } dx \Bigg] dy = \int_{-R}^{+R} \int_{-\sqrt{R^{2}-y^{2}}}^{\sqrt{R^{2}-y^{2}}} \Big[(-2qy+B_{back})(2q) dx \Big] dy
\end{equation}
\begin{equation}
    = \int_{-R}^{+R}  \Bigg[ \int_{-\sqrt{R^{2}-y^{2}}}^{\sqrt{R^{2}-y^{2}}}  \Bigg(-4q^{2}y+2qB_{s}\exp \Big(\dfrac{-y}{2H_{B}}\Big) \Bigg) dx \Bigg] dy.
\end{equation}
\noindent
Integrating this, using  again the modified Bessel function in  (\ref{eq:ModifiedBesselFunction_FirstOrder}) (with $t=y/R$ and $z=R/2H_{B}$), gives 
\begin{equation} \label{eq:NetVerticalTensionForce}
    F_{tension,y} = 8\pi qRB_{s}H_{B}I_{1}(\tau).
\end{equation}
Equation (\ref{eq:NetVerticalTensionForce}) confirms that the net internal tension force is zero when the large-scale background field is not present ($B_{s}=0$). Moreover, it is now clear that the introduction of background field ($B_s \ne 0$) creates a net internal tension force that can either support the rise of the flux tube ($F_{tension,y} > 0$) or hinder its rise ($F_{tension,y} < 0$) based on the sign of the product of twist and background field. This is commensurate with our earlier findings from the simulations that it is the alignment of the twist and the background field that dictate the behaviour.  If $q$ and $B_s$ have the same sign (i.e. the azimuthal field of the tube and the background field are aligned at the bottom of the tube), then the product $q B_s$ is positive and the tension force is upwards, supporting rise.  Note that for a fixed $B_{s}$, the net tension force varies linearly with $q$.

The third part of the force budget that we need to consider is the tension force induced in the large-scale background field by the rise of the structure upwards into this field, which we denote by $F_{back,y}$.  This opposes the rise of the tube. To estimate $F_{back,y}$, we examine the tension force induced in a rise that proceeds to a general height, $H$. For a successful rise in our simulations, $H$ will be approximately equal to or greater than the maximum vertical size of the simulation box, $y_{max}=4$. Tension forces in curved field lines can be written as square of the magnitude of the field divided by the radius of curvature, pointing inwards to the curve. If we assume that a rising tube lifts all of the field it encounters during the rise and concentrates it into a narrow region wrapped around the tube, then a description of the net tension force (in the $y$-translated coordinate system used earlier) opposing the vertical rise induced during the rise might be reasonably given by the integrated square amplitude of the field divided by the radius of the tube:

\begin{equation}
    F_{back,y} = -\dfrac{1}{R} \int_{0}^{H} B_{back}^{2} ~ dy
\end{equation}

\begin{equation}
    = -\dfrac{B_{s}^{2}}{R} \int_{0}^{H} \exp \Big(\dfrac{-y}{H_{B}} \Big) dy
\end{equation}

\begin{equation}
    =-\dfrac{H_{B}B_{s}^{2}}{R} \Bigg(1-\exp\Big(\dfrac{-H}{H_{B}}\Big) \Bigg).
\end{equation}
For small $H$, i.e. $H<<H_{B}$, we can rewrite the exponential function using series expansion and keep only terms up to first order, giving 
\begin{equation}
    F_{back,y} = -\dfrac{HB_{s}^{2}}{R}.
\end{equation}
Clearly then the tension force induced in the background field is linear in $H$ and will be negligible when the height to which the flux tube rises, $H$, is very small, as expected. When the flux tube rises more significantly, i.e.  $H>>H_{B}$, the net induced tension force tends to a constant value
\begin{equation} \label{eq:InducedTensionForce}
    F_{back,y} = \dfrac{-H_{B}B_{s}^{2}}{R}.
\end{equation}
The tension therefore only varies rapidly over a height of about a scale height $H_B$ and its role can be bounded above nicely by equation (\ref{eq:InducedTensionForce}). 

If we make the assumption that the internal buoyancy and tension forces in the tube-like concentration do not change much from their initial values, then the marginal case that defines the transition from a failed rise to a successful rise is given by a balance of the three elements described above, i.e. 
\begin{equation} \label{eq:MarginalCase_Initial}
    F_{tension,y} + F_{buoyancy,y} + F_{back,y} = 0.
\end{equation}
Assuming some non-negligible rise, $H >> H_{B}$,  using Equations (\ref{eq:NetBuoyancyForce_QuadraticQ}),  (\ref{eq:InducedTensionForce}) and (\ref{eq:NetVerticalTensionForce}), this becomes

\begin{equation}\label{eq:MarginalCase_FinalEquation}
    \Bigg[\dfrac{2\theta (m+1) \pi R^{4}}{T_{avg}} \Bigg] q^{2} + \Bigg[8\pi RB_{s}H_{B}I_{1}(\tau)\Bigg] q  + \Bigg[\dfrac{\theta (m+1)\pi R^{2}}{2T_{avg}}+\pi B_{s}^{2}H_{B}^{2}RI_{1}(2\tau) - \dfrac{H_{B}B_{s}^{2}}{R}\Bigg] = 0.
\end{equation}

\noindent
This equation is quadratic in $q$ and we can calculate its roots $(q_1,q_2)$ for given $H_{B}$ and $B_{s}$ (given also $\theta$, $m$, $T_{avg}$ and $R$ which are assumed to have fixed values throughout this paper). These roots $(q_1,q_2)$ are the model estimates of the $(q_-,q_+)$ from the simulation data in Figure  \ref{fig:q_Bs_SurveyChart}
earlier.  

\begin{table}
    \centering
    \begin{tabular}{|c|c|c|c|}
    \hline
    $\mathbf{B_{s}}$ &   $\mathbf{H_{B}}$ & $\mathbf{q_{1}}$ &   $\mathbf{q_{2}}$ \\
    \hline
     \multirow{3}{*}{0.03} &  0.075   &   -3.14 + 2.26i   &   -3.14 - 2.26i   \\
    &   0.125   &   -2.97 + 2.32i   &   -2.97 - 2.32i   \\
    &   0.175   &   -2.93 + 2.23i   &   -2.93 - 2.23i   \\
    \hline
    \multirow{3}{*}{0.05}  &    0.075   &   -9.01   &   -1.45   \\
    &   0.125   &   -8.62   &   -1.29   \\
    &   0.175   &   -8.71   &   -1.05   \\
    \hline
    \multirow{3}{*}{0.10}   &   0.075   &   -20.72  &   -0.21   \\
    &   0.125   &   -19.99  &   0.17    \\
    &   0.175   &   -20.08  &   0.56    \\
    \hline
    \multirow{3}{*}{0.16}   &   0.075   &   -33.90  &   0.41   \\
    &   0.125   &   -32.75  &   1.04    \\
    &   0.175   &   -32.87  &   1.64    \\
    \hline
    \end{tabular}
    \caption{Roots of the quadratic equation \ref{eq:MarginalCase_FinalEquation} for some example values of $B_{s}$ and $H_{B}$}
    \label{tab:Quadratic_Roots}
\end{table}

\begin{figure}
    \centering
    \includegraphics[width=16cm, height=6cm]{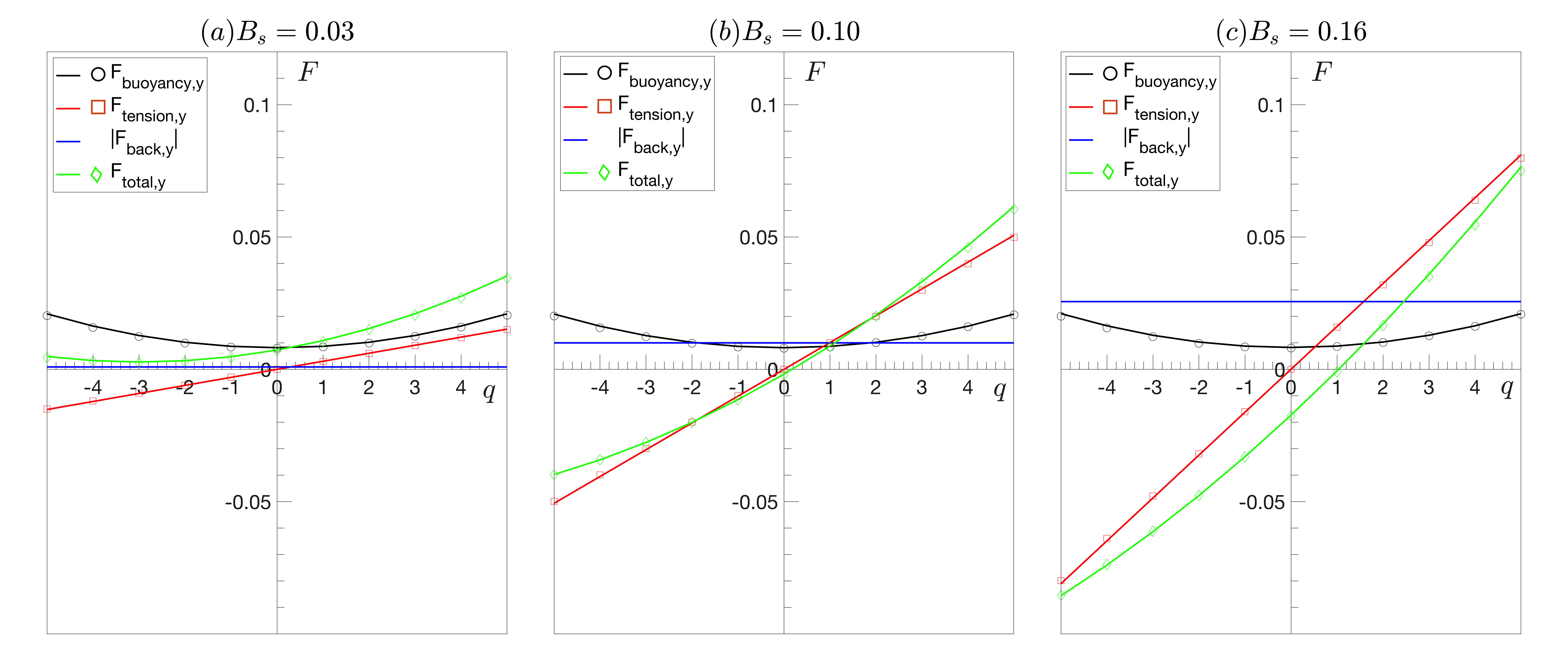}
    \caption{Shown are forces acting on the flux tube in vertical direction calculated from the theoretical model: Area-integrated net buoyancy and tension forces internal to the tube;  the upper bound estimation of the opposing tension force induced in the background field by the rise; and,  the total upwards vertical force (the sum of internal buoyancy, internal tension and induced external tension).  These are  plotted as a function of the flux tube twist, $q$, for $B_{s}= (a)~0.03,~(b)~0.10$ and $(c)~0.16$, for the canonical value of  $H_{B}=0.125$. The symbols show the corresponding integrated forces calculated using the simulation data.}
    \label{fig:AllForces_AreaIntegrated_Analytical}
\end{figure}

\begin{figure}
    \centering
    \includegraphics[width=11cm, height=12cm]{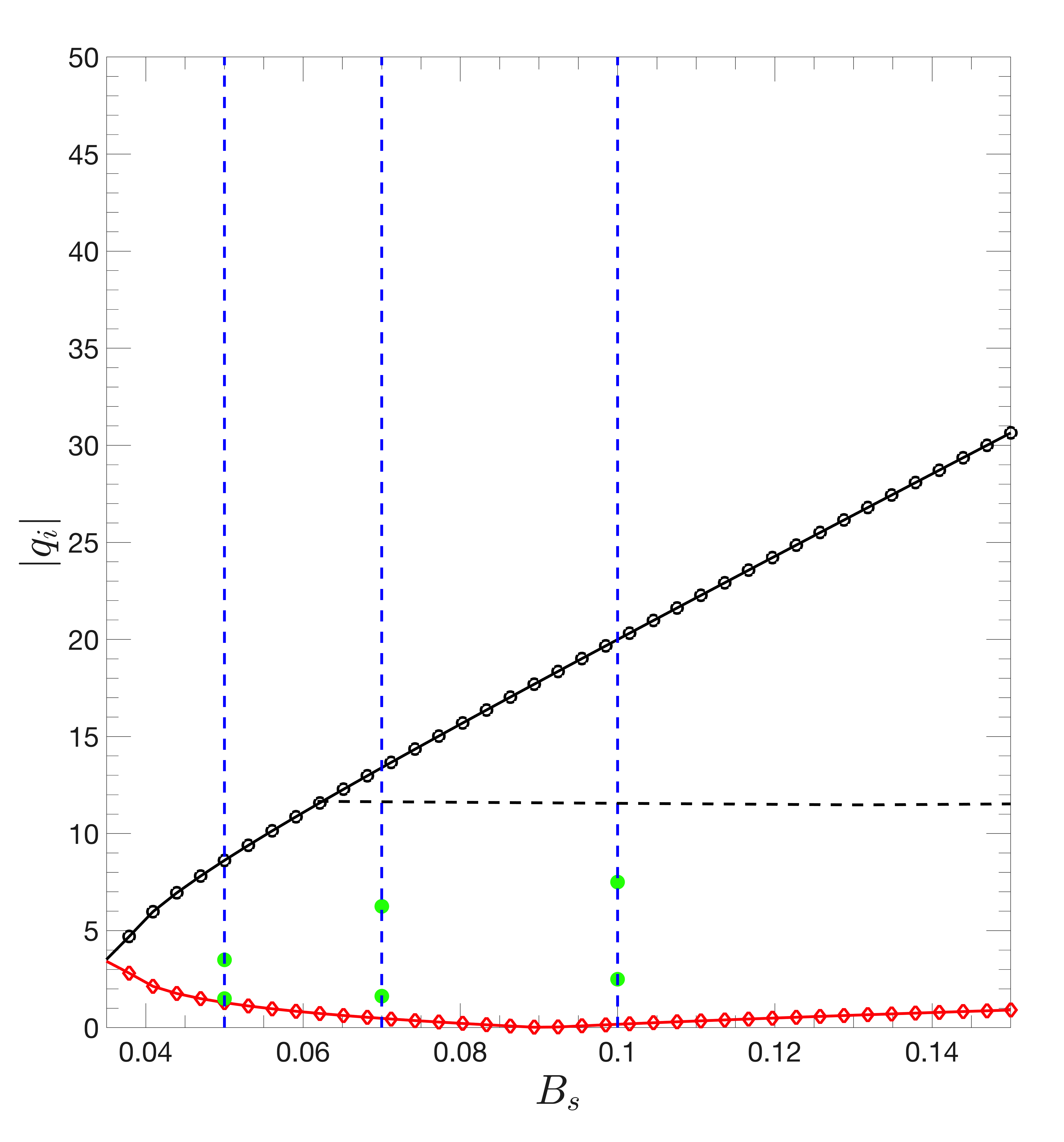}
    \caption{The SRR given by the model as a function of $B_{s}$ for $H_{B}=0.125$. Black circles and red diamonds represent the absolute value of roots of Equation (\ref{eq:MarginalCase_FinalEquation}).  For these cases with positive $B_s$, $q_1$ is negative and is the root with larger modulus (black circles).  The other root $q_2$ has smaller modulus (red diamonds) and changes sign from positive to negative at around $B_s \sim 0.09$.  Blue dashed vertical lines in the plot are the cases studied using numerical simulations and the black dashed line represents the highest allowable twist that keeps the thermodynamic conditions inside the flux concentration physical (i.e. such that the density remains positive).  The green circles are the limits of the SRR found from the numerical simulations.}
    \label{fig:Hb125_SRR_Explained}
\end{figure}

Table \ref{tab:Quadratic_Roots} shows examples of the numerical values of these roots for some selected $B_{s}$ and $H_{B}$.  Figure \ref{fig:AllForces_AreaIntegrated_Analytical} accompanies the table and shows the relative contributions of the individual components of magnetic buoyancy, internal tension and background field tension forces, plus their total, from the model equations (\ref{eq:NetBuoyancyForce_QuadraticQ}), (\ref{eq:NetVerticalTensionForce}) and (\ref{eq:InducedTensionForce}) respectively, as a function of $q$ for the canonical background field configuration with $H_{B}=0.125$ and the $B_s$ indicated.  We will use these cases as illustrative examples.   A more complete evaluation of the roots of the canonical case is given later in Figure  \ref{fig:Hb125_SRR_Explained}, and other $H_{B}$ values are addressed in the next section.
In Table \ref{tab:Quadratic_Roots}, we see that for very weak values of $B_{s}$ $(B_{s} \le 0.03 ~\text{at}~ H_{B}=0.125)$, there exists no real roots for $q$. This is consistent with the fact that all values of twists, whether positive or negative, will rise successfully when only a very weak background field is present (assuming that $q$ is above the minimum threshold necessary to maintain coherency of the structure)\footnote{This is likely similar to the $q \gtrsim 0.25$ threshold for the  $B_{s}=0$ case, but note that the presence of background field may adjust this.}.  Figure \ref{fig:AllForces_AreaIntegrated_Analytical}a illustrates this result, as $F_{total,y}$, the sum of the forces on the left hand side of equation (\ref{eq:MarginalCase_FinalEquation}), does not cross the horizontal axis for any $q$ at this $B_s$. For intermediate and high values of $B_{s}$, we get two real roots and these are asymmetric in $q$.  For example, we find that the roots for $B_{s}=0.10$ are $q=(q_{1},q_{2}) \approx (-19.99,~0.17)$ and for $B_{s}=0.16$ are $q=(q_{1},q_{2}) \approx (-32.75,~1.04)$. This asymmetry in the roots defines the model definition of the SRR. For example, if $q_{1} < 0$ and $q_{2} > 0$, then rise occurs for structures with $q<q_{1}$ and $q>q_{2}$ and there is a region of $|q|$ between $|q_{1}|$ and $|q_{2}|$ where structures possessing $+|q|$ and $-|q|$ behave differently ($+|q|$ rises, $-|q|$ does not). The canonical examples from Paper I for $B_{s}=0.10$, showed that $q=2.5$ rises but $q=-2.5$ did not, and so $|q|=2.5$ lies in the SRR of this $B_s$, which we can see here is given by $0.17 < |q| < 19.99$. More generally, in these examples with positive $B_s$, positively-twisted flux tube that are fairly weak can rise whereas  negatively-twisted flux structures of the same strength (or even relatively much stronger) cannot. 

Figure \ref{fig:AllForces_AreaIntegrated_Analytical} clarifies the origin of the asymmetrical roots and therefore the SRR in terms of the individual force contributions. The buoyancy force (black lines derived from Equation (\ref{eq:NetBuoyancyForce_QuadraticQ}) with associated circles from simulation data -- see later) is quadratic in $q$ and can now be seen to be dominated by the contribution from the initial density perturbation due to the magnetic structure (the second term on the right hand side of Equation (\ref{eq:NetBuoyancyForce_QuadraticQ})), since the buoyancy force is very close to  symmetric in $q$ and only very weakly dependent on $B_{s}$. The asymmetry in the $q$ dependence of the total upwards force on the tube (green line; diamonds) instead stems from the net vertical internal tension force in the tube (red line; squares). This is linear in $q$, and significant in size because it depends on $B_{s}$ and $H_{B}$ rather than these quantities squared. For the $B_s>0$ shown here, negative $q$ creates a negative tension force serving to reduce the effectiveness of the buoyancy force, retarding the  rise of negative $q$ structures. This effect can completely inhibit the rise unless the buoyancy force (proportional to $q^{2}$) is large enough to compensate. Where $q>0$, the internal tension force is positive and aids in the rise of the structure.  The tension in the background field (blue line; no symbols) is independent of $q$ and only serves to reduce the chances of rise evenly across the range of $q$ at fixed $B_s$ (but is dependent on $B_{s}^{2}$).  The net force, given by the left hand side of equation (\ref{eq:MarginalCase_FinalEquation}) (green line; diamonds) is therefore a shifted and tilted quadratic function.  Where this function passes through zero are the roots $q_{1}$ and $q_{2}$, and now it is clear that the effect of the background field on the internal tension of the tube is responsible for introducing the major asymmetry into this function and therefore is the culprit for creating the asymmetric roots leading to the existence of a SRR. 

The results have been discussed here only for $B_s>0$.  The results for $B_s<0$ can be inferred by the symmetry in the $q-B_s$ space, as noted earlier with relation to the simulation data in Fig. \ref{fig:q_Bs_SurveyChart}.  For example, in an equivalent of Fig. \ref{fig:AllForces_AreaIntegrated_Analytical} for $B_s<0$, the buoyancy force and the tension force induced in the background field would remain essentially the same, but the internal tube tension force would switch sign. The roles of $q_1$ and $q_2$ would therefore be reversed, with the asymmetry such that $|q_1| << |q_2|$. 
A general definition of the model SRR such that it is valid for any orientation of $B_{back}$ in terms of the roots of the quadratic is given similarly to Equation \ref{eq:GoldilocksZone_data} by
\begin{equation} \label{eq:GoldilocksZone}
    \text{SRR}_{\text{model}} = \{q ~~{\text s.t.} ~~|q| \in [\text{min}(|q_{1}|,|q_{2}|),~\text{max}(|q_{1}|,|q_{2}|)]\}
\end{equation}
 
Figure  \ref{fig:Hb125_SRR_Explained} shows an example of the delineation of the SRR in $q-B_s$ space via this method (at the canonical parameter values, in particular $H_B=0.125$; other $H_B$ are done shortly).
The red diamond points show $\text{min}(|q_{1}|,|q_{2}|)$ values and the black circular points show $\text{max}(|q_{1}|,|q_{2}|)$.  For these cases with $B_s>0$, the root of smaller modulus is $q_2$.  The SRR is the region between the lines. Structures with values of $q$ of either sign such that $|q|$ is above the upper limit of the SRR (line of black points) all rise.  For $B_s \gtrsim 0.09$, where $q_1$ and $q_2$ are of opposite sign,  structures with both signs of $q$ would fail to rise if $|q|$ is below the lower limit (red diamonds).  For $0<B_s \lesssim 0.09$, where $q_1$ and $q_2$ are both negative, both signs would rise for $|q|$ below the red line.  Only within the SRR range do structures of one sign rise whereas structures with the opposite sign do not.  Positive twist rises preferentially for the cases shown here with $B_s>0$, but negative twist would rise preferentially if $B_s<0$. The SRR definition given in Equation  (\ref{eq:GoldilocksZone}) is presented as a range of $q$ at fixed $B_s$ and is therefore a function of $B_s$.  The SRR is actually a region in $q-B_s$ space and could equally well be presented as a range of $B_s$ that is a function of $q$, as is clear from both the data and the model, as shown in Figs. \ref{fig:q_Bs_SurveyChart} and \ref{fig:Hb125_SRR_Explained}.

The model proposed therefore seems to explain well the nature of the dynamics that have been observed.  However, a number of simplifications were introduced, namely:  the assumption that the background temperature variation across the tube was reasonably represented by its average; that it is reasonable to omit the $q$ dependent term in equation (\ref{eq:NetBuoyancyForceExpanded}); and, that it is a good approximation to estimate the internal tube forces (buoyancy and internal tension) only at $t=0$ and then use that to deduce the subsequent $t>0$ rise behavior.  Since we have all the data from the simulations, it is constructive to compare what we can in the model with the numerical simulations in order to verify our assumptions. We therefore start by crudely calculating the net internal forces in the structure at $t=0$ by summing over the relevant quantities inside the tube area as follows:
\begin{equation}
    F_{buoyancy,y,sim} =   \sum_{i,j~s.t.~r_{i,j} \le R} \Bigg[-\dfrac{\partial}{\partial y} \Big(P_{gas}+\dfrac{B^{2}}{2} \Big)+ \rho g \Bigg]_{i,j} \Delta x_i \Delta y_j
\end{equation}
\begin{equation}
    F_{tension,y,sim} = \sum_{i,j~s.t.~r_{i,j} \le R} \Big[ \left( {\bf B} \cdot \nabla \right) {B_y} \Big]_{i,j} \Delta x_i \Delta y_j
\end{equation}
\begin{equation}
    F_{total,y,sim}=F_{buoyancy,y,sim} + F_{tension,y,sim} + F_{back,y},
\end{equation}
where $r_{i,j}=x_i^{2}+(y_j-y_{c})^{2}$ with $i,j$ the indices of the discrete simulation grid in the $x$ and $y$ direction respectively. For each relevant simulation that we have performed, we plot the values of these net internal forces as the appropriately-colored symbols in Figure \ref{fig:AllForces_AreaIntegrated_Analytical}. Internal magnetic tension (red squares) calculated using the simulation data exactly matches with the model estimate.  The black circles representing the internal tube buoyancy from the simulations do have a slight linear dependence on $q$ added to the symmetrical quadratic term, but so slight that it is not visually apparent in the figure, vindicating our omission of this term in the model. 

The major source of uncertainty in our model then comes from the estimate of $F_{back,y}$.  This quantity is a very rough approximation to one aspect of more complex dynamics. Omitted effects are the deformed geometry of the structure, effects of the wake and the overlaying field on the wake, any drainage from the main structure, and diffusive effects, at least. These other effects are hard to model and would be hard to check against data even if they were modeled, since they are not compactly spatially located to the structure and are time-evolving. The discrepancy associated with these errors shows up in comparisons of  the roots that we obtain by solving the quadratic Equation (\ref{eq:MarginalCase_FinalEquation}) and the SRR range deduced from observing the simulations.  The upper and lower bounds of the SRR deduced from the simulations are shown for a few sample $B_s$ as the green circles in Figure \ref{fig:Hb125_SRR_Explained}. Although the simulation SRR boundaries are only visually extracted from a coarse set of simulations and are therefore relatively inaccurate, it is clear from this comparison that the simulated SRR is substantially narrower than the model predicts. It appears that our estimate of the rise-quenching effects from the background field is indeed quite a severe underestimation, as we imagined. We could attempt ad hoc changes to the model (e.g. increasing the radius of curvature of the retarding external tension force in the background field to reduce its effect and to account for the topology of the rising structure later in its rise), or try to calibrate the model to the data, but since the simulations are fairly coarsely spaced in parameter ($q-B_s$ space) and the boundaries between successful and failed rise are sometimes not entirely obvious, then this seems unlikely to be particularly accurate.  Furthermore, the omitted effects are probably dependent on the parameters (such as $B_s$ and $H_B$) too, and therefore a simple calibration is unlikely to work universally.  However, even with some significant  uncertainty, the model still provides a good explanation of the dynamics behind the SRR, and gives useful predictions of the trends seen in the simulations, at least.  In particular, it appears to be a  robust result that the extent of the SRR grows and the root of lower modulus grows (in signed value) with increasing $B_s$ (see also Fig. \ref{fig:Width_SRR_WeakHb_Bs_Marker} later).  These facts have a significant influence on the expected distribution of the signs of the emerging helicity, as will be discussed in detail in Section \ref{subsec:MultipleMonteCarlo}.

\begin{figure}
    \centering
    \includegraphics[width=8cm, height=10cm]{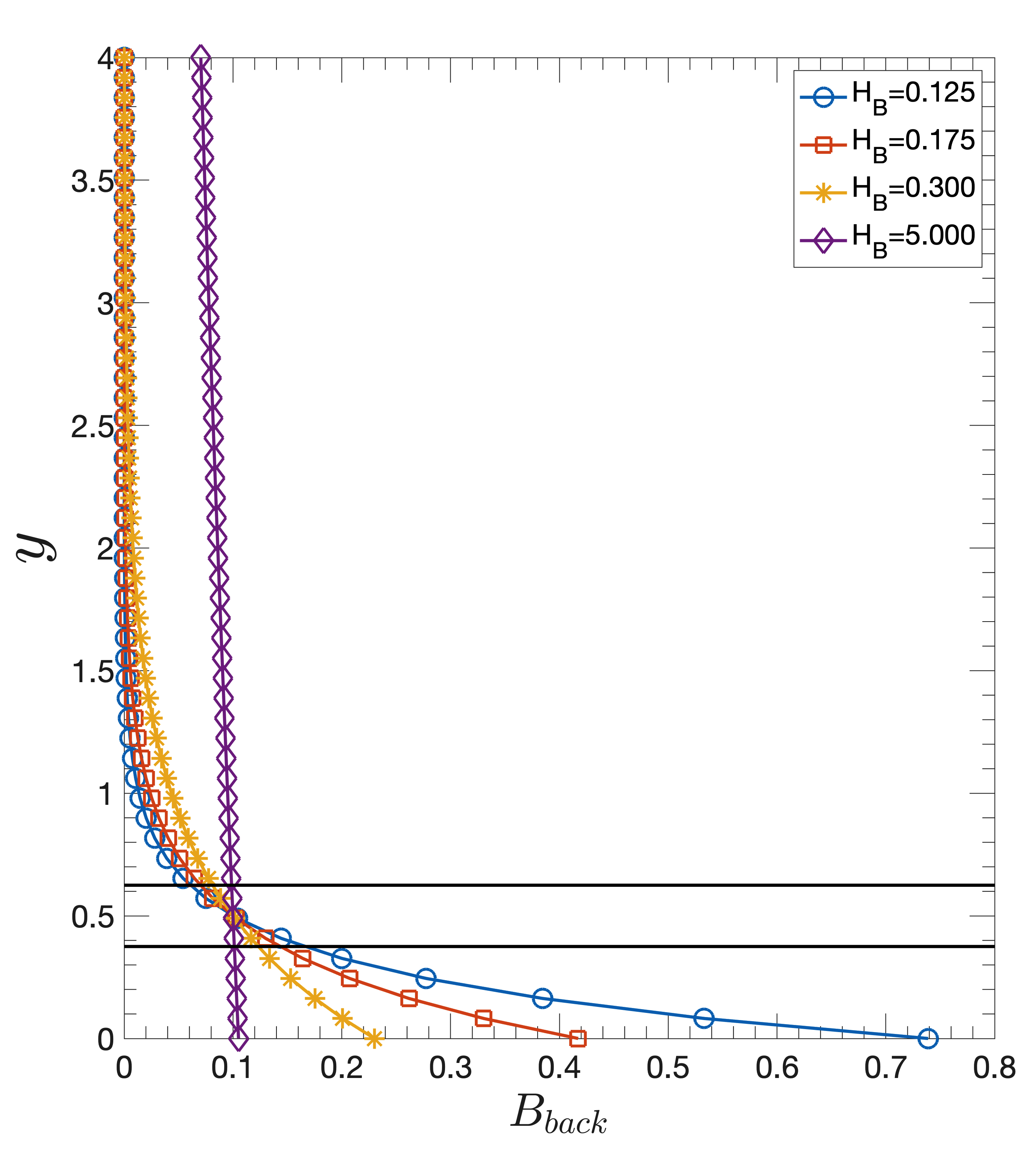}
    \caption{Variation of background field as a function of height for different scale heights $H_B$ for $B_{s}=0.1$. Dotted horizontal lines indicate the vertical location of the flux tube.}
    \label{fig:BackgroundField_HbVariation}
\end{figure}

\begin{figure}
	\includegraphics[width=\columnwidth, height=18 cm]{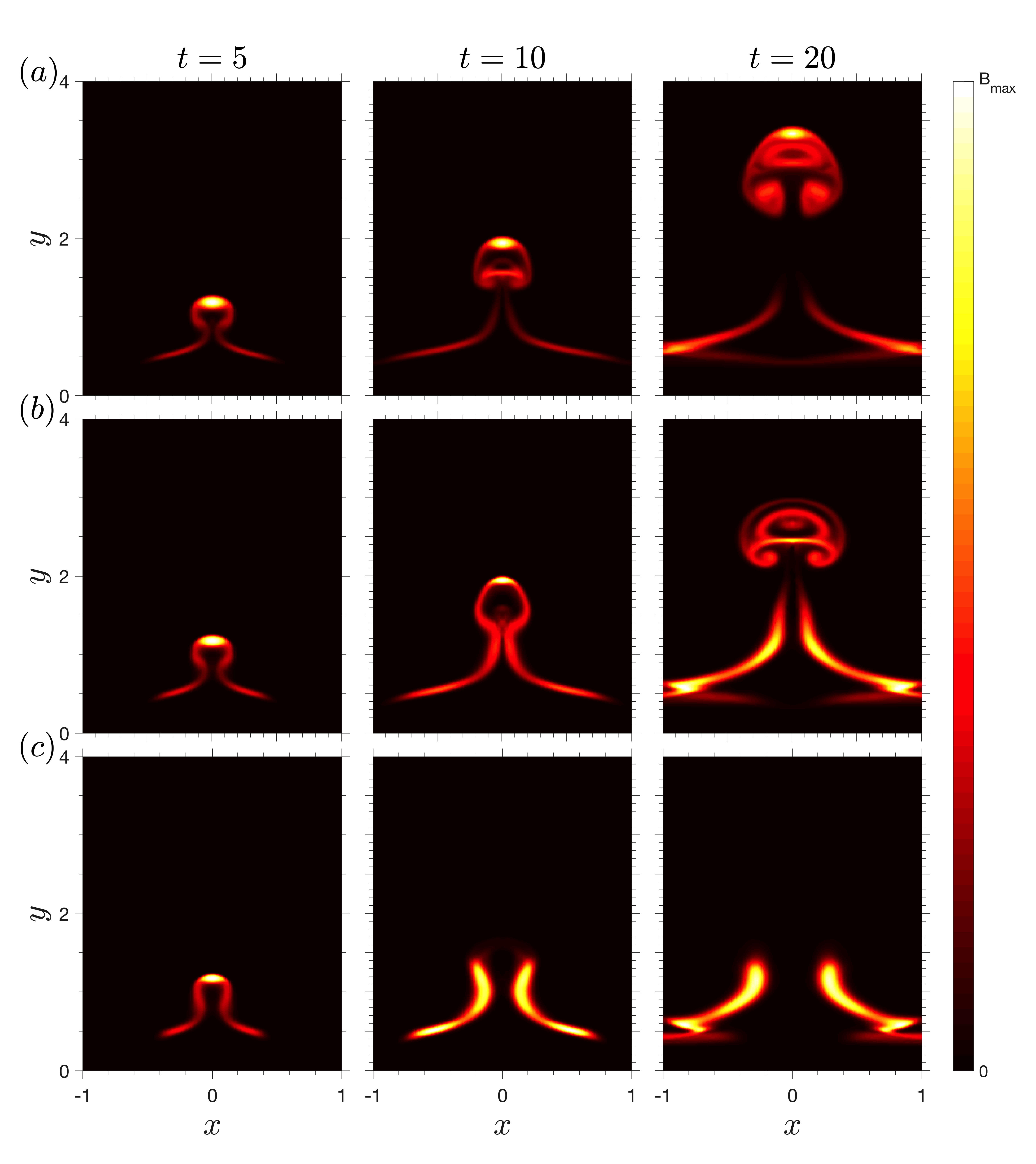}
    \caption{Effect of the background field scale-height variation on the dynamics. Intensity plots of normalized $B_{z}(x,y,t)$ for $H_{B}=0.125,~0.175,~0.300$ (a, b and c respectively) when $B_{s}=0.10$ and $q=2.5$ in all cases. All the subplots are scaled with respective max($B_{z}$).}
    \label{fig:Hb_variation_075_125_175}
\end{figure}

\begin{figure}
    \centering
    \includegraphics[width=\columnwidth, height=12cm]{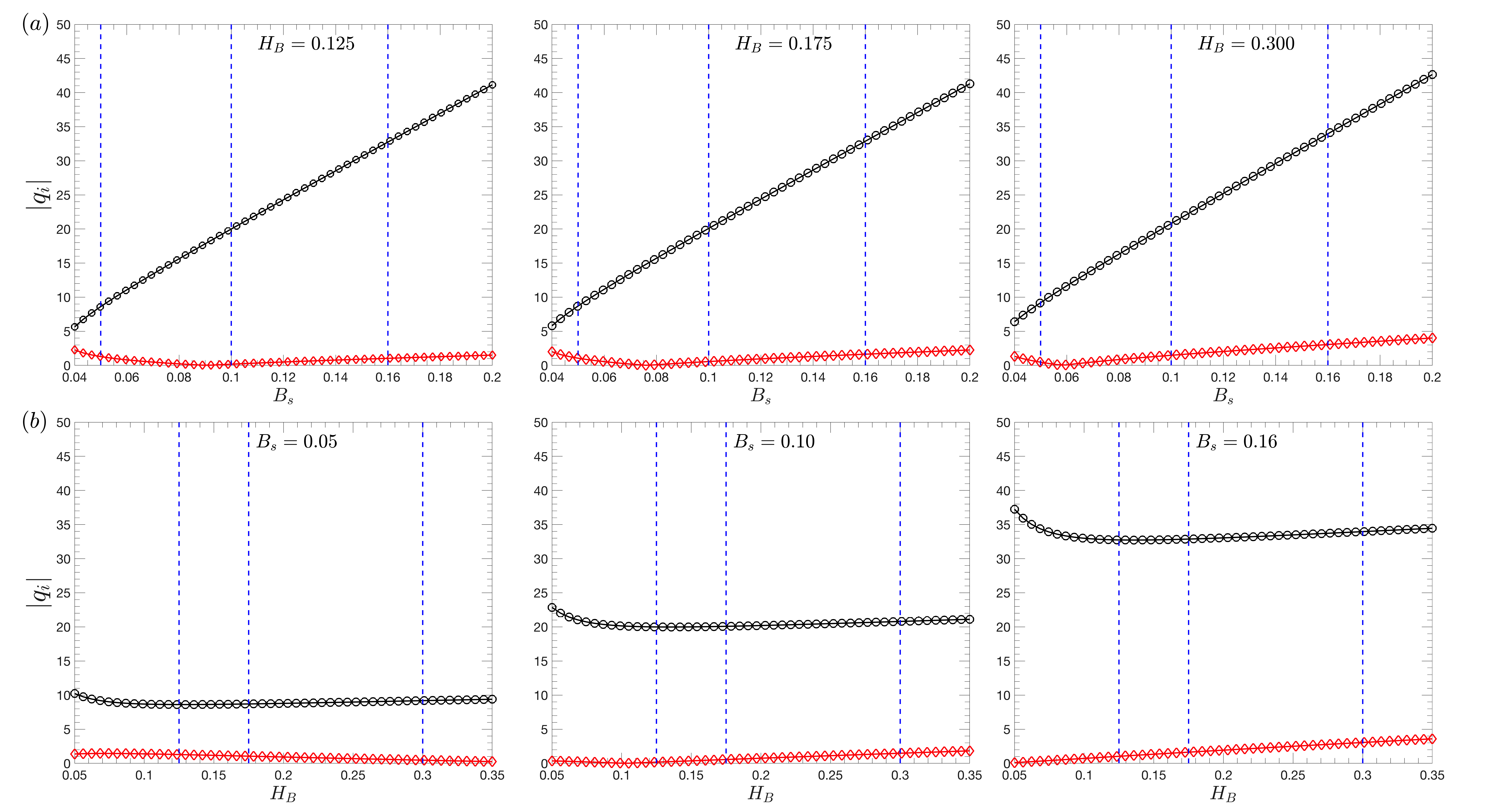}
    \caption{Absolute value of the roots of Equation (\ref{eq:MarginalCase_FinalEquation}). The region between the black circles and red diamonds is the SRR. (a) Variation of SRR for a fixed $H_{B}$ as a function of $B_{s}$. Blue vertical dashed lines correspond to $B_{s}=0.05,~0.10,~0.16$, which are values where we have comparable simulations. (b) Variation of SRR for a fixed $B_{s}$ as a function of $H_{B}$, Blue vertical dashed lines correspond to $H_{B}=0.125,~0.175,~0.300$, which are values where we have comparable simulations.}
    \label{fig:GoldilocksZone_Combined}
\end{figure}

\begin{figure}
    \centering
    \includegraphics[width=\columnwidth]{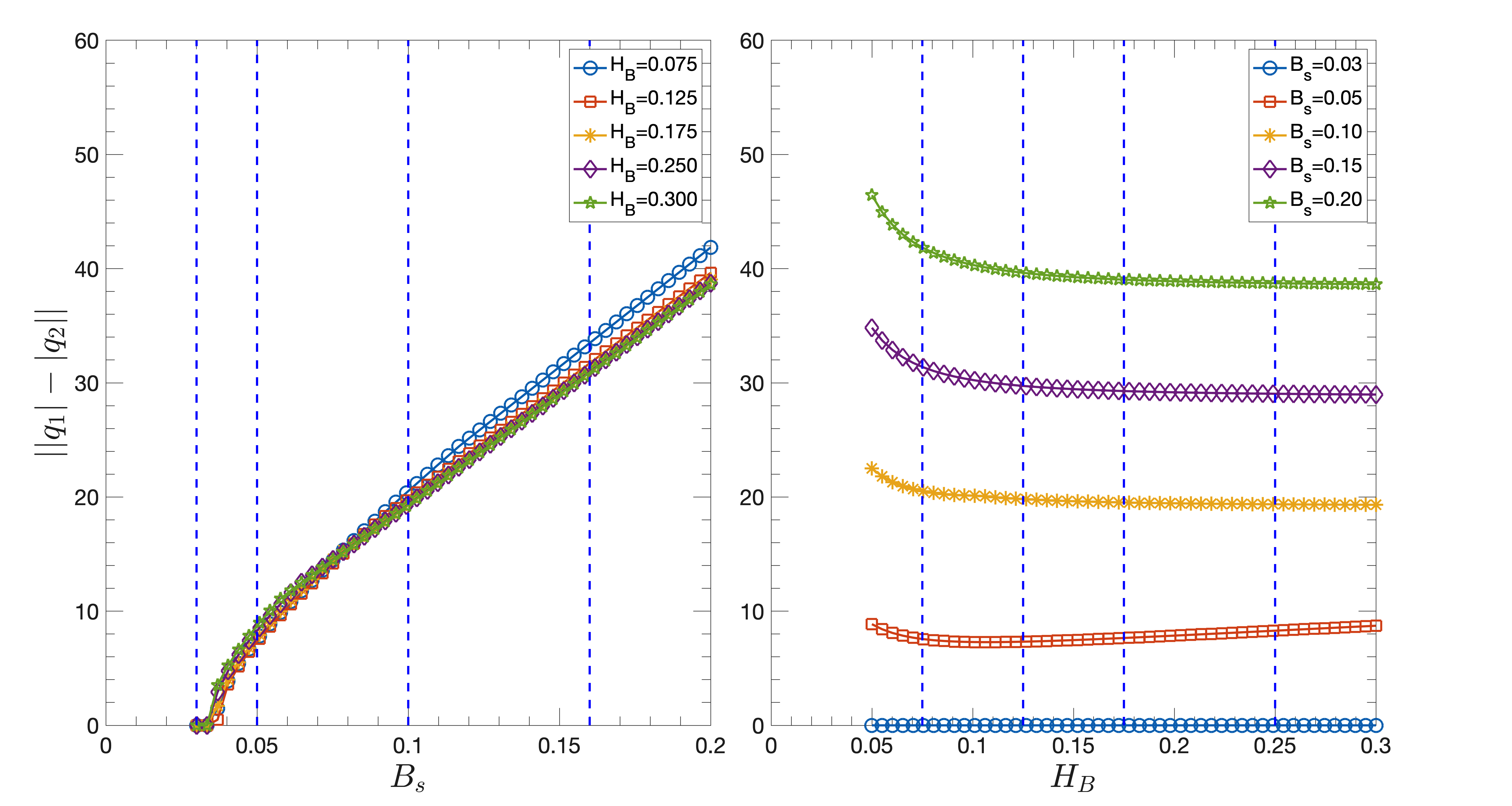}
    \caption{Width of the SRR, $|\left(|q_{1}|-|q_{2}|\right)|$ plotted as a function of (a) $B_{s}$ for different $H_{B}$ and (b) $H_{B}$ for different $B_{s}$.}
    \label{fig:Width_SRR_WeakHb_Bs_Marker}
\end{figure}

\subsection{Dependence on the scale-height of the background field} \label{sec:ScaleHeight}

We have now established the importance of the presence of the large-scale background field on the dynamics of a rising flux tube, concentrating so far on the relationship between the background field strength and the twist strength, and their relative orientations. We now examine the dependence on the configuration of the background field, which we have characterized as an exponential with scale height $H_{B}$.  Figure \ref{fig:BackgroundField_HbVariation} shows the variation of the background field, $B_{Back}$, as a function of height, $y$, for some example values of the scale-height, $H_{B}$, when $B_{s}=0.1$. The dotted horizontal lines in the figure show the initial vertical location of the flux concentration. Recall that $B_s$ specifies the field strength at the center of the tube so all these lines intersect at that point with value $B_s=0.1$.  This figure is intended to emphasize the fact that, when $H_B$ is small, the background field variation across the flux tube is substantial, whereas for high values of $H_B$ the background field changes very little with height overall and hence its variation across the tube is small.  Since this differential is key to the new dynamics discussed above, we might expect $H_B$ to play a role.

Figure \ref{fig:Hb_variation_075_125_175} shows intensity plots of the (normalized) axial field $B_{z}$ at three different times in the evolution of system for three values of scale height, $H_{B}=0.125,~0.175,~0.300$, whilst keeping the background field strength and twist of the flux tube constant (at $B_{s}=0.10$ and $q=2.5$). Overall, Figure \ref{fig:Hb_variation_075_125_175} shows that the rise of the flux structure is quenched for higher $H_{B}$. Figure \ref{fig:Hb_variation_075_125_175}$a$, exhibiting the evolution for the canonical $H_{B}=0.125$, shows that a typical flux structure -- a ``squashed'' head with trailing vortices -- rises through the background field towards the top of the simulation domain. Figure \ref{fig:Hb_variation_075_125_175}$b$, for $H_{B}=0.175$, shows another overall successful rise, but this time there is considerable drainage of the axial flux $B_{z}$ away from the tube head down the the overlaying background fieldlines, plus some significant change to the geometry of the head and the trailing vortices.   Note that the flux tube is further from the top of the simulation domain at $t=20$ in Fig.~\ref{fig:Hb_variation_075_125_175}$b$ and is therefore rising more slowly compared to the previous case, even though it appears to be still rising as a fairly independent structure. Figure \ref{fig:Hb_variation_075_125_175}$c$ shows the case with an even larger scale height, $H_B=0.3$. In this case, there does not appear to be any coherent rising flux structure. 

The quenching of the rise of the flux structure with the increase in scale height of the background field can be understood from a physical perspective, with confirmation from our mathematical model. One can imagine that increasing the scale height of the background field could lead to three effects that could potentially affect the dynamics. Firstly, a larger $H_B$ redistributes the background field in the tube and this affects the magnetic pressure that drives the rise of the tube.  The analytical approximation in Equation (\ref{eq:NetBuoyancyForce_QuadraticQ}) reveals that this dependence is quite complex, being proportional to $H_B^2 I_1(R/H_B)$.  This is a function that increases with $H_B$ and therefore its contribution to the magnetic buoyancy is enhanced at larger $H_B$.  However, as we discussed earlier, this magnetic pressure contribution is much smaller than the contribution from the initial density perturbation in the buoyancy term proportional to $g=\theta(m+1)$.  This first effect is therefore likely not very significant.  

The second potential effect is that an increase in $H_B$ leads to a decrease in the differential of the amplitude of the field across the tube vertically.  It is this differential that drives the selection mechanism via its effect on the magnetic tension in the tube.  Any effect (either increased likelihood of rise when $B_s$ and $q$ are of the same sign, or decreased likelihood when $B_s$ and $q$ are of opposite signs) is reduced for larger $H_B$.  In the case shown in Figure \ref{fig:Hb_variation_075_125_175}, we might expect the tension-driven enhancement of the rise to be reduced by larger $H_B$. This expectation can again be examined with the analytical model, this time via Equation  (\ref{eq:NetVerticalTensionForce}).  There, it can be seen that the direction of the force depends on the sign of $q B_s$ and also that the amplitude again has a complicated dependence on $H_B$: $F_{tension,y} \sim H_B I_1(R/2H_B)$.  This dependence is almost constant,  only very slightly decreasing over a significant range of $H_B$.  This second effect is therefore again probably not very important.

The third effect remains as the likely major dynamical factor.  Increasing $H_B$ also serves to increase the integrated field that contributes to the retarding force of the overlaying background field.  One might expect larger $H_B$ to imply a larger value of the integrated field and therefore a large force countering the rise.  Since this is an integral over the distance travelled upwards by the tube, this is probably a substantial effect.   These ideas can again be verified by the analytical model, where this time the dependence can be seen from Equation (\ref{eq:InducedTensionForce}) to be straightforwardly  linear: $F_{back,y} \sim H_B$.  As mentioned previously, this estimate is also potentially an underestimate, with some significant factors related to other effects omitted.  This also lends some credence to this being the dominant physical term here, although we do not know the $H_B$ dependence of the other effects.

Having established some physical intuition, we can further employ the analytical model developed in Section \ref{sec:ForceAnalysis} to analyze the broader dependence of the SRR on the scale height of the background field, $H_B$. Figure \ref{fig:GoldilocksZone_Combined} plots the SRR from Equation (\ref{eq:GoldilocksZone}) in $q-B_s$ space for fixed $H_{b}$ (Figure \ref{fig:GoldilocksZone_Combined}$a$) and in $q-H_B$ space for fixed $B_{s}$ (Figure \ref{fig:GoldilocksZone_Combined}$b$).  The blue vertical dashed lines in each of the subplots show the values where numerical simulations exist for comparison with the model.  For the values of $H_{B}$ surveyed, by comparing the panels of \ref{fig:GoldilocksZone_Combined}$a$ or examining an  individual panel of Figure \ref{fig:GoldilocksZone_Combined}$b$ at fixed $B_s$, we see that the demarcation (and therefore the extent) of the SRR (in $|q|$) only varies a very moderate amount.  This is commensurate with our earlier conclusion that the tension effect that drives the existence of the SRR is only very weakly dependent on $H_B$. On the other hand, individual panels of Figure \ref{fig:GoldilocksZone_Combined}$a$ or comparison of the panels in  Figure \ref{fig:GoldilocksZone_Combined}$b$ confirms  that varying the background field strength, $B_{s}$, can significantly change the extent of the SRR for fixed values of $H_B$.  This is all again clarified by the analytical model where the tube internal tension term in Equation  (\ref{eq:NetVerticalTensionForce}) that establishes the SRR is linearly dependent on $B_s$, but is roughly constant with $H_B$ (via the term $H_B I_1(R/2H_B)$). An important, and perhaps unexpected, result overall here is that an SRR still exists even for high $H_B$, although the $q$ values at which it occurs become less physical.

It now becomes necessary to reconcile the above statements with the dramatic quenching of the rise of a structure with increasing $H_B$ observed in Figure \ref{fig:Hb_variation_075_125_175}.
First, it should be noted again that we expect realistic values of the twist $q$ to be of order $q < 10$.  Higher values create unphysical (negative) values of the density (see horizontal dashed line in Fig. ~\ref{fig:Hb125_SRR_Explained}) and other work has shown that high $q$ are unstable. For instance, the initial Gaussian profile for the axial field used in Section \ref{sec:PreviousResults} would have a critical value of $q_{cr} = 1/R = 8$ for the flux tube to be kink unstable in three dimensions \citep{Linton:Longcope:Fisher:1996}. This restriction means that realistic flux structures probably lie much closer to the lower limit of the SRR than the upper. If $q$ is such that it is close to the lower boundary, then even a small variation in $H_B$ and a weak dependence of the SRR boundary on $H_B$ can have a dramatic effect. This is essentially what was demonstrated through the simulations shown in Figure \ref{fig:Hb_variation_075_125_175}. The simulations in this figure are at  $q=2.5$ and $B_s=0.10$, corresponding to the second blue dashed line in Figure \ref{fig:GoldilocksZone_Combined}$a$ and the middle panel in Figure \ref{fig:GoldilocksZone_Combined}$b$.  In the latter, for $q=2.5$, the moderate upwards tilt of the lower boundary of the SRR (red diamonds) with increasing $H_B$ shifts the solution from inside the SRR (for $H_B=0.125$) to below the SRR (for $H_B=0.3$), thereby switching it from a rising solution to a quenched solution.  A small change in $H_B$ can therefore have an effect due to the proximity of natural solutions to the lower boundary of the SRR.

To emphasize the above dependencies, Figure \ref{fig:Width_SRR_WeakHb_Bs_Marker} plots the width of the SRR, $||q_{1}|-|q_{2}||$, as a function of (a) $B_{s}$, for a range of $H_{B}$, and, (b) $H_{B}$, for a wide range of $B_{s}$ (as in the previous figure). These figures solidify our interpretation that the width of the SRR is only very weakly dependent on $H_{B}$ but significantly dependent on $B_s$.  Note again that at very low values of $B_s$  (e.g.  $B_{s}=0.03$) there is no SRR since the differential tension force required to generate the necessary asymmetry for the SRR is negligible.

\subsection{Multiple flux concentrations} \label{subsec:MultipleMonteCarlo}

With an understanding of the dynamics of a selection mechanism established for a single flux concentration, in this section we examine the collective effect that leads to the observations that are known as the SHHR. The SHHR observations accumulate the signs and strengths of the helicity of emerging active regions over the whole solar cycle (and eventually multiple cycles) to extract any net bias.  The SHHR observations have a large ``scatter", in the sense that not all individual active regions agree with the rules -- overall, only 60-80\% of active regions appear to concur -- and the degree of agreement appears to vary over the solar cycle (\citet{Seehafer:1990, Pevtsov:Canfield:Metcalf:1995,Abramenko:Wang:Yurchishin:1997, Bao:Zhang:1998, Bao:Ai:Zhang:2000, Bao:Pevtsov:Wang:Zhang:2000, Hagino:Sakurai:2005, Hao:Zhang:2011, Singh:Kaplya:Brandenburg:2018}). The scatter and its temporal variation is not well understood. In this section, we therefore attempt to use our model to  create synthetic SHHR observations with a view to potentially elucidating reasons for these effects.

To reproduce SHHR observations, we simulate the evolution of multiple (toroidal) flux structures through different background (poloidal) field strengths representing different parts of the solar cycle.  We fix the values of the other parameters at the canonical values: $m=1.5,~H_{B}=0.125,~R=0.125,~\theta=0.25$.  For each field strength, we perform a series of Monte Carlo (MC) simulations where, in each simulation, a wide domain is initialised with a number of flux tubes with randomly assigned twist strength (both positive and negative) and horizontal location.  We then examine whether the selection effects shown for single flux tubes leads to a persistent bias over the whole cycle (simulated by our range of background field strengths). It may seem likely that this is the case as we have identified a selection mechanism, but this needs to be confirmed, since this setup allows the possibility of new dynamics emerging from the interaction of tubes, and the randomness may potentially swamp any bias.

In order to have reasonably significant statistics for each case in this MC study, we perform fifty independent simulations (each containing multiple tubes) for each background field strength. Since we are not addressing the origin of the flux structures, our assumption is that they are formed from a sheet of toroidal field, likely by a magnetic buoyancy instability (see for e.g. \citet{Vasil:Brummell:2008,Guerrero:Kaplya:2011}).  If this were to be the case, we might then expect multiple flux structures to be formed at the same time in a turbulent environment, and such structures would most likely possess random strengths, twists and separations drawn from some distribution. We therefore choose each of our simulations in the MC series for a particular background field strength to start with five flux concentrations with random initial twists $(-5 \le q \le 5)$ and random distance, $d$, between them (with the constraint, $d \geq 2R$, so that they do not overlap), embedded initially at the same (canonical) height in the prescribed overlying large-scale background field. The random values are drawn from a uniform distribution over the range given.  
Our setup certainly has some significant degree of arbitrariness because of the limitations in our quantitative understanding of the amount of twist in flux tubes, the distance between them and their strengths, at their origination in the deep solar interior. For example, our choice of drawing the random values of twist from a uniform distribution might not be very realistic and a Gaussian with a larger range might be better (or perhaps even a skewed distribution to account for the effects in the formation of tubes not included in our model, e.g. rotation:  see \cite{Chatterjee:Mitra:Rheinhardt:Brandenburg:2011}).
However, as mentioned earlier, there are some constraints on the value of $|q|$. Firstly, $|q| \gtrsim 12$ (the dashed line in Fig. \ref{fig:Hb125_SRR_Explained}, which is actually a very weakly decreasing function of $B_s$) is unphysical, since then the azimuthal field is sufficiently strong induce a negative density in the structure.  Secondly, structures with $|q| \gtrsim 8$ are unstable to kink instabilities (according to \citet{Linton:Longcope:Fisher:1996}, which does not fit our model exactly, but is likely a reasonable guide).
Therefore, our chosen distribution (uniform, $|q| \le 5$) seems reasonable, in that it makes it likely that the initial simulation setup has a mixture of weak and reasonably strong concentrations of both helicities, and, if anything, overemphasizes the stronger twists (that are more likely to violate the SHHR). An MC series of this type (fifty simulations, five random tubes each) is reported for four different background field strengths, ranging from relatively weak to relatively influential ($0.01 \le B_s \le 0.10$). We identify the weak background field with the beginning and end of the cycle and the stronger background with the peak of the cycle, although we have no direct observational evidence from the deep solar interior of the relative strengths of these fields from which to set their values.

\begin{figure}

\includegraphics[width=\columnwidth]{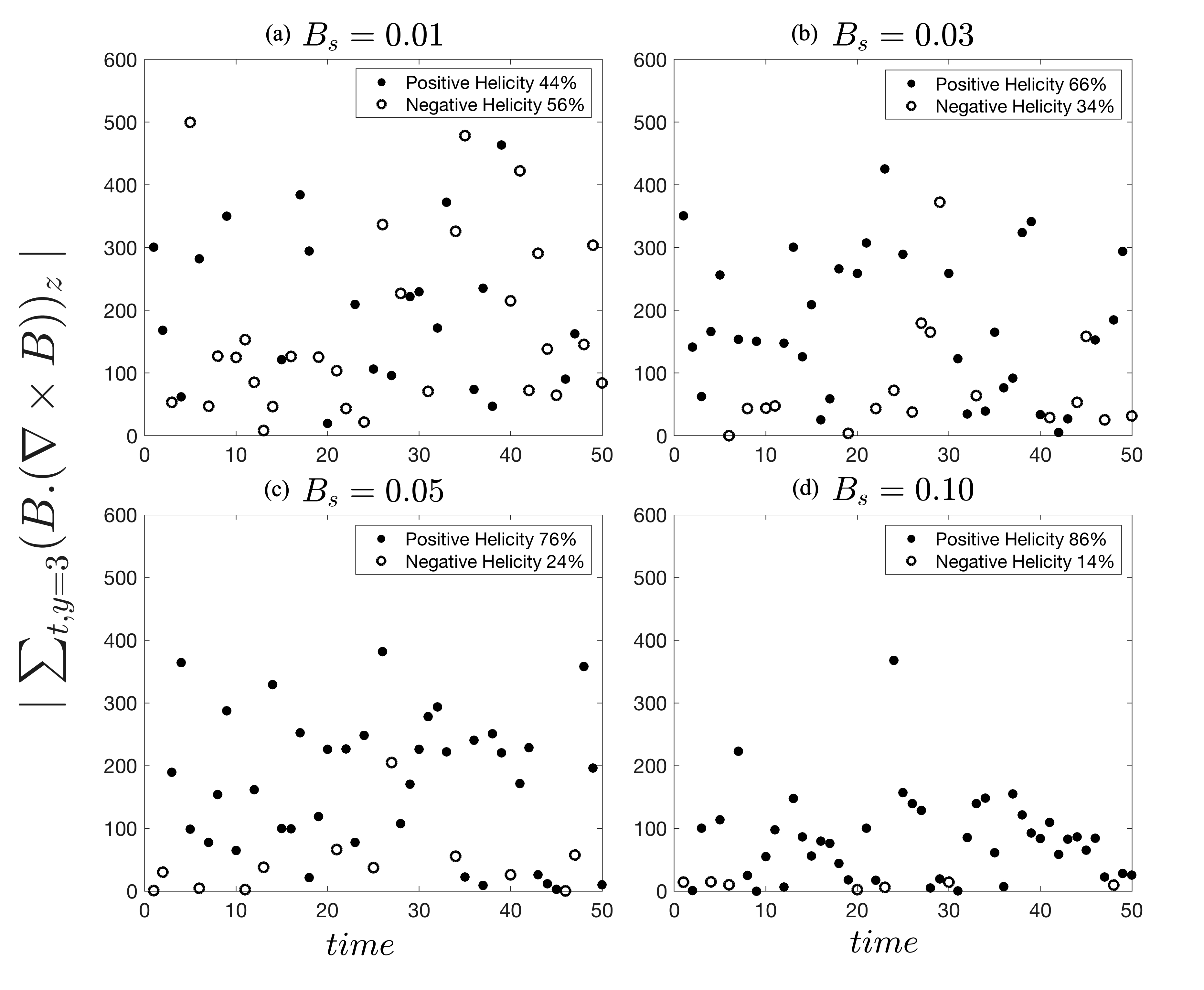}
\caption{Synthetic SHHR.  Each dot shows the sign (solid circle = positive; open circle = negative) and aggregated strength (vertical axis) of the horizontal component of the current helicity that passes through $y=3$ during a the interval $t=0$ to $t=70$ from a simulation involving five flux structures embedded within the background field of strength $B_s$ shown for each panel.  The flux structures in each simulation have randomly assigned twist and separation from a distribution.  Each panel contains 50 of such MC simulations, and represents a particular time in the solar cycle, with low $B_s$ representing times around solar minimum, and higher $B_s$ being associated with solar maximum.  Percentages show the relative distribution of positive and negative helicity.  Positive helicity is expected to dominate for the configuration.}
    \label{fig:SyntheticSHHR_figure}
    
\end{figure}

Fig.~\ref{fig:SyntheticSHHR_figure} shows the results of the MC simulations for each of the four different background field strengths. Each circle in the plot represents an accumulation over a specified time interval $[t_0,t_1]$ in the simulation of the absolute value of the z-component of the net horizontal current helicity measured along a line at $y=3$.  We call this quantity $H_3$:
\begin{equation}
H_3 = \left|\int_{t_0}^{t_1} \int_{-1}^{1}[B.(\nabla \times B)]_{z} ~dx ~dt \right| = \left| \int_{t_0}^{t_1}\int_{-1}^{1} B_{z}\Big( \dfrac{\partial B_{y}}{\partial x} - \dfrac{\partial B_{x}}{\partial y} \Big) ~dx ~dt \
\right|.
\label{eqn:h_3}
\end{equation}
This quantity is evaluated from the simulations as a sum over the $x$ grid points at height $y=3$ at a particular time, and over the data output in the time interval from $t_{0}=0$ to $t_{1}=70$. 
This quantity represents the magnitude of the net helicity of all the flux concentrations that successfully rise through $75\%$ of the model solar interior in the simulation setup. The open and filled circles indicate the net sign, corresponding to net negative and positive helicity respectively. Each panel in the plot shows the 50 MC simulations for a given field strength, $B_{s}$, representing a given time in the solar cycle, plotted against the simulation number, which could be loosely interpreted as short time intervals at that point in the cycle,  but are in reality just different realizations whose ensemble average becomes meaningful. The overall bias in the emerging helicity at any specific time in the solar cycle is therefore given by any statistically-significant deviation of the percentage of each sign of helicity away from a $50\%-50\%$ balance. These percentage values are noted on the plots. Note that owing to the orientation of the tube's axial field along the positive z direction and the background field along the the positive x direction, our setup corresponds to the the southern hemisphere in the first half of the cycle as shown in Fig. 5a, where positive helicity is expected to dominate by the SHHR. Furthermore, to relate to observations, we are assuming that some three dimensional process leads to arching of our simulated flux concentrations for emergence so that the axial helicity that we measure horizontally in 2D is strongly correlated with the vertical helicity component that emerges in the observations.  We will test these assumptions in 3D in the future.

Fig.~\ref{fig:SyntheticSHHR_figure} displays a systematic trend in the relative percentages of the two signs of accumulated helicity with increasing background field strength. For a very weak background field strength of $B_{s}=0.01$ in Figure \ref{fig:SyntheticSHHR_figure}a, we find that the surviving and accumulating helicity at $y=3$ is roughly equally divided amongst both signs of helicity.
There is a slight bias towards an anti-SHHR result (more negative helicity) as has actually been observed in the solar case, but we do not dwell on this as this seems fairly statistically  insignificant here, since the number of MC simulations is only 50.  At higher background field strengths (which we relate to closer to solar maximum in the the solar cycle), we find results that are definitely commensurate with the SHHR. At $B_{s}=0.03$, we find that around $66\%$ of the successfully rising flux concentrations possess positive helicity; at $B_{s}=0.05$ we find that $78\%$ are positive; and at $B_{s}=0.1$, $86\%$ are positive. These results are in agreement with the overall SHHR trend, which is that positive helicity should generally be preferred in this Southern hemisphere setup (and, of course, the results for the Northern hemisphere can be inferred by symmetry).
Furthermore, these results also agree with the observed trend that there are fewer violations to the SHHR (i.e. better agreement) around solar maximum and more violations (worse agreement) around solar minimum \citep{Hagino:Sakurai:2005}. Note that we correlate a stronger deep interior background poloidal field in our model with solar maximum and a weaker poloidal field with solar minima. Since we do not know the deep interior fields, we rely upon mean-field dynamo models to justify this \cite[see e.g.][and references there-in]{Charbonneau:2020}. Many such models exhibit that the poloidal and toroidal fields are reasonably in phase, and the latter are widely used as a proxy for the butterfly diagram.

\begin{figure*}
	\includegraphics[width=\textwidth, height=9 cm]{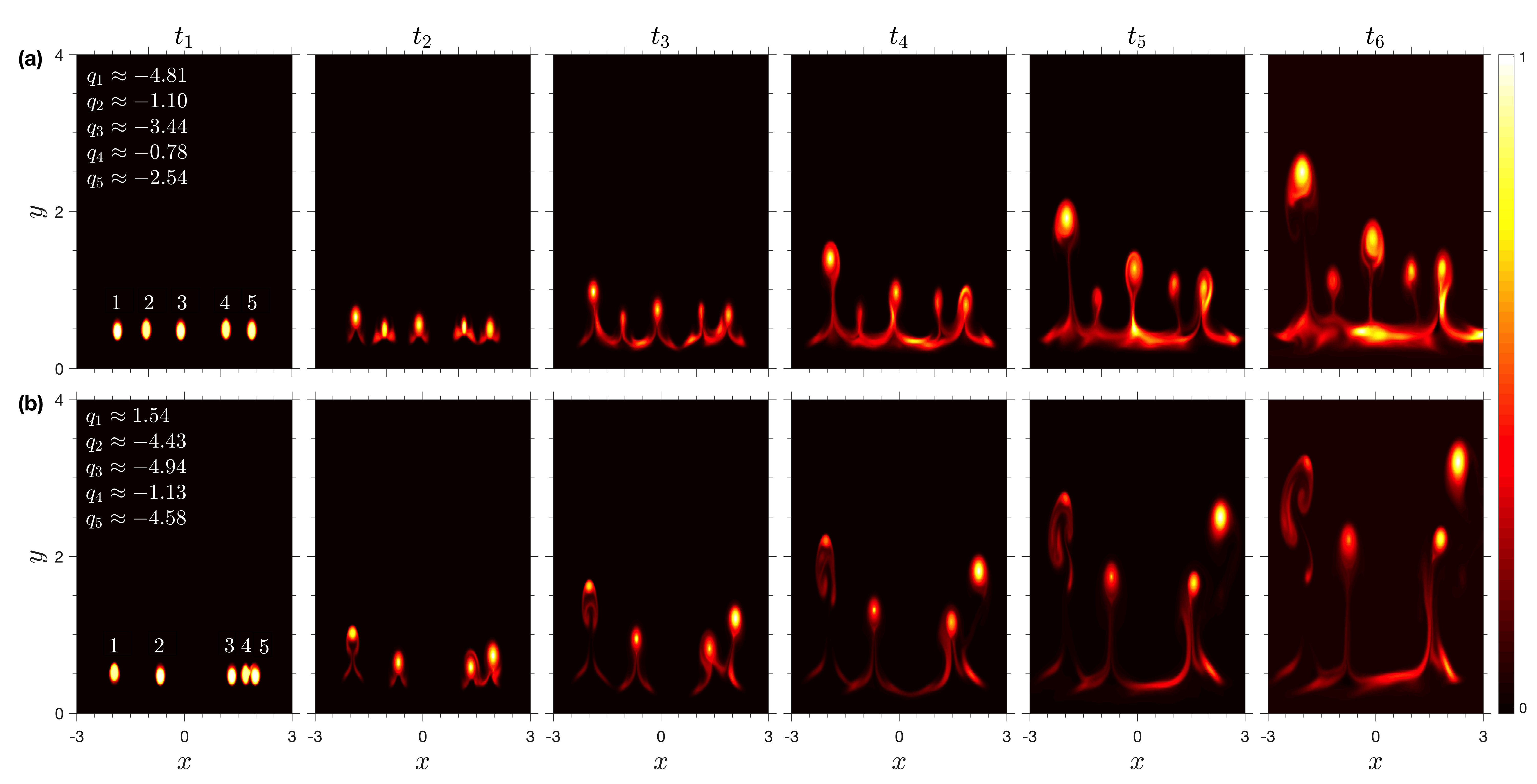}
    \caption{Solar Hemispheric Helicity rule violations.  Both panels show intensity plots of normalized $B_{z}(x,y,t)$ at $(t_{1},t_{2},t_{3},t_{4},t_{5},t_{6})=(1,11,21,31,41,51)$. Panels a and b are independent instances of the MC simulations at $B_{s}=0.05$, chosen to depict particular SHHR violation mechanisms. With the axial field, $B_{z}$, out of the plane and poloidal background field aligned in the positive x-direction, this setup is representative of the solar southern hemisphere (see Fig.~\ref{fig:3D_Cartoon_SHHR}a). The sign and strength of the field line twist ($q$) at $t_{1}$ is indicated  for all the concentrations, with the subscript corresponding to the numbering of the concentrations along the x-axis. Panel (a) shows violations occurring due to the rise of strongly twisted tube with negative twist. Panel (b) shows a violation occurring due to the interaction of two negatively twisted tubes.}
    \label{fig:Helicity_Violation_Figure}
\end{figure*}

These major results can be interpreted fairly straightforwardly in the light of the work in Sections \ref{sec:AnalyticalResults} and \ref{sec:ForceAnalysis}, and, in particular, the $q-B_s$ diagrams from simulations (e.g. Fig. \ref{fig:q_Bs_SurveyChart}) and the model (e.g. Figs.
\ref{fig:Hb125_SRR_Explained},
\ref{fig:GoldilocksZone_Combined} and see also Fig. \ref{fig:Width_SRR_WeakHb_Bs_Marker}).
In general, these plots show that there is only a narrow SRR at weak $B_{s}$ but that the SRR grows wider in extent in $|q|$ as $B_{s}$ increases (a robust trend, even though the data and the model do not agree precisely on location of the boundaries of the SRR due to the inaccuracies of the model, as discussed earlier).
Violations of the SRR would mainly be detected when the twist of a structure is strong enough for it to lie above the SRR (although this is not the only possibility, as discussed shortly).  Since there are good physical reasons for the value of $|q|$ to be capped above, clearly, where the SRR is narrower at lower $B_s$, more violations might be expected to occur, since our distribution of $q$ in the MC simulations is more likely to place them above the SRR region (and below the cap).  In the limit of the SRR vanishing, all structures lie outside the SRR and either don't rise, or are equally likely to rise, therefore yielding percentage values for the helicity sign distribution close to 50\%-50\%.  As the SRR becomes wider at higher $B_s$, the number of violations drops as $q$ values drawn from the distribution become less likely to lie above the SRR. This leads to more complete compliance with the SRR and  a ratio of the emerging helicity signs that would tend towards 100\%-0\% in favor of the sign of helicity expected from the SHHR (positive in our example cases).  Indeed, for higher $B_s$ ($B_s \gtrsim 0.06$ in Fig. \ref{fig:Hb125_SRR_Explained}), the cap of our distribution ($|q|=5$) lies below the upper bound of the SRR from both the model and data, and almost 100\% agreement would be expected, potentially violated only by very low $|q|$ lying in the small region below the lower bound of the SRR (red diamonds in the model). Structures with such sub-SRR twist are only expected to rise when $B_s \lesssim 0.09$ (where both roots $q_1$ and $q_2$ are negative); violations are not expected for sub-SRR values of twist for $B_s \gtrsim 0.09$ since both signs should not rise, but interestingly in this region another method of violation (discussed shortly) operates, reducing the agreement with the SHHR from 100\%. Regardless of these details, the trend to fewer violations for stronger $B_s$ seems very robust and explainable by the selection mechanism found here.

\begin{figure*}
	\includegraphics[width=\textwidth, height=5 cm]{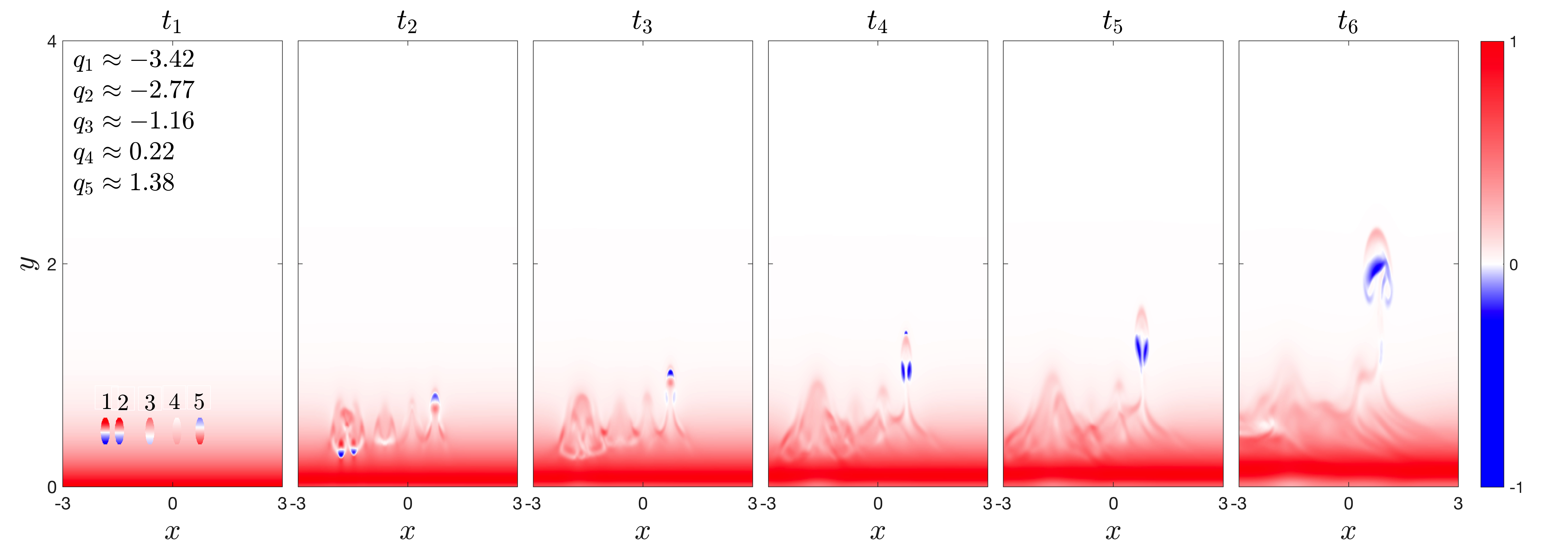}
    \caption{Violation of the SHHR via restructuring of the twist.  Panels show normalized $B_{x}(x,y,t)$ for $B_{s}=0.1$ at $(t_{1},t_{2},t_{3},t_{4},t_{5},t_{6})=(0,5,9,15,20,40)$. Tube 5 has anti-clockwise (positive) twist at $t=t_{1}$ but gets rearranged to a clockwise (negative) twist by $t=t_{5}$.  This mechanism is quite common for strong background fields and weakly twisted structures.}
    \label{fig:Bx_Restructure}
\end{figure*}

The above result, related to the expected degree of scatter in the observations and its dependence on the solar cycle, is a reflection of how violations to the SHHR are achieved in the MC simulations.  We therefore investigate the causes of these violations in more details now. To exhibit the mechanisms we find, we focus on the cases with $B_{s}=0.05$ and $0.10$.  We find three distinct, relevant reasons for violations which are discussed below:

\begin{enumerate}
    \item Strongly twisted flux tube: The first, and main, reason, mentioned previously, is simply the existence of strongly twisted  flux concentrations of the ``wrong" sign (where, here, ``wrong" means that the sign of helicity of the structure violates the SHHR; in the cases shown here, the ``wrong" sign is negative since the SHHR would predict predominantly positive helicity). If the structures are sufficiently strongly twisted to lie above the SRR, then these structures will rise. As proof of this from the simulations, Figures~\ref{fig:Helicity_Violation_Figure}$a,b$ show examples of the time evolution of normalized axial magnetic field strength, $B_{z}$, from the MC simulations performed with $B_{s}=0.05$. The twist sign and strength of individual flux concentrations are shown, numbered by their relative position along the  horizontal axis at $t=t_{1}$.  Fig.~\ref{fig:Helicity_Violation_Figure}$a$ shows a case where the randomly-assigned twists of all the flux concentrations in the simulation are negative (clockwise) and therefore not what would be expected from the SHHR. However, tube 1 has almost the maximum twist allowable from our distribution and we see that at $t=t_{6}$, this tube has risen significantly through the model solar interior and is approaching the $y=3$ level where we measure the net helicity.   Note that tubes 3 and 5 are behaving similarly, but to a lesser extent.  From Figures \ref{fig:q_Bs_SurveyChart} and \ref{fig:Hb125_SRR_Explained} it can be seen that these high values of $|q|$ lie outside the numerically-identified SRR (but inside the potentially inaccurate upper boundary of the model SRR) and therefore might be expected to rise. This is the main source of scatter, and was predicted from our single structure simulations.
    
    \item Tube-tube interaction: Fig.~\ref{fig:Helicity_Violation_Figure}$b$ exhibits the second class of violation, which is a case where interactions between multiple structures do play an important role. Of the five flux concentrations in this figure, four of them have $q<0$ and the remaining one ($q_{1}$) has $q>0$. The first three, $q_1,q_2, q_3$, behave as one might expect: $q_1$ is moderately strong and positive and therefore from being in the SRR, it is expected to rise; $q_2$ and $q_3$ are strong but negative and therefore may still rise since they are above the SRR (by the previous method of violation).  The other two, $q_4$ and $q_5$, are very close together, thanks to the use of random distances between concentrations in these MC simulations. The dynamics of the tubes then are no longer those of individual single tubes as interactions between structures become important. Since both these concentrations have the same twist, they coalesce very quickly and form a single, more strongly buoyant region.  This structure is still negative but has effectively more twist and more buoyancy, and therefore rises more quickly, becoming a violation of the previous kind due to its strong effective twist.  This shows that separation of the structures can be an important element of the dynamics, as the original twist distribution alone would not have predicted the resulting outcome.
    
    \item Flux tube twist rearrangement: A third scenario that can lead to violations of the SHHR is particularly peculiar and results from the conversion of weak ``correct" signed structures into the ``wrong" sign during the rise. An example is shown from one of the MC simulations for the case of $B_{s}=0.1$ in Fig.~\ref{fig:Bx_Restructure}.  These types of violations are quite common at higher $B_s$ and account for the majority of the violations shown in  Fig.~\ref{fig:SyntheticSHHR_figure}$d$.  Fig.~\ref{fig:Bx_Restructure} shows the evolution of $B_{x}$, a clear indicator of twist of the tube, in this example. The first three of the structures initially have negative twist, and the remaining two are  positive. The first three tubes do not rise successfully since they have only moderate twist strength lying in the SRR and are therefore filtered out by the selection mechanism.  Tube 4 has positive twist but is too weak to rise (lies below the SRR). Tube 5 is the peculiar case.  This structure has an intermediate positive twist and is expected to rise from consideration of the SRR. However, this structure undergoes a topological arrangement early in its transit, flipping the sign of its twist (visible in the plot as a flip from blue over red in the $B_x$ field to red over blue).  The exact mechanism for the flip is a little difficult to discern, but appears to be facilitated by the interaction of the strong background field wrapping around into the trailing vortices of the original structure. The structure continues to rise despite this rearrangement, and therefore emerges with the ``wrong" sign.

\end{enumerate}

Overall, these MC simulations have revealed the perhaps expected result, that the SRR mechanism can be responsible for the aggregated helicity bias that is the observed SHHR, but has further identified good reasons for the scatter in the observations and its temporal variation, some even beyond explanation by our concept of the existence of an SRR.

\section{Discussion and Conclusions}

In this work, we have examined the rise of a tube-like concentration of magnetic flux in the presence of a large-scale background field.  Our model is aimed at elucidating the processes which bring magnetic structures from the deep solar interior towards the solar surface, where observations are taken.  The core of our work has been to relax some of the constraints of previous models.  Much intuition has been gleaned from vastly simplified models using either the thin flux tube approximation or finite cross-sectional tubes, but in general, these studies have examined the evolution of preconceived cylindrical magnetic structures that are isolated, in the sense that they are embedded in a field-free background.  These simulations do not address the origin of such structures.  Studies of magnetic buoyancy instabilities that do try to examine this question show that the resultant structures are concentrations of strong magnetic field amongst a volume-filling, weaker, large-scale background field.  Our premise has been that the dynamics of concentrations might be substantially different from isolated structures.  At this stage, we have examined the most obvious extension of previous models, where we embed the preconceived tube-like structures from before in a large-scale background field, to convert it from an isolated structure into a concentration.  We continue to ignore the origin of these fields and the dynamics of convection in the rise of the concentration, leaving these added complexities for later studies.

We indeed succeed in showing that relaxing these assumptions can lead to drastically different rise dynamics of the magnetic flux structure. In particular, our results reveal that the rise of the flux structure is (perhaps not surprisingly) impeded by the background field, although unexpectedly it can be completely quenched by a relatively weak background field (a factor of 10-20 weaker than the peak tube strength). This quenching is achieved by a combination of (i) tension induced when stretching out the overlaying field as it is lifted by the rising structure, (ii) a drainage of the axial flux that supplies the buoyancy of the structure out along the overlaying background field lines, and, (iii) suppression of the trailing vortices that self-propel the structure after the initial buoyant rise. 

Much more surprising is that the strength of the background field ($B_{s}$) required to quench the rise of the magnetic structure is dependent on the relative orientation of the twist of the structure ($q$) and the background field.  When the background field is aligned with the azimuthal field ($B_\theta$) at the bottom of the structure, a net tension force is created inside the tube that is directed upwards, complementing the buoyancy force driving the rise.  When the background field is aligned with azimuthal field at the top of the structure, the net internal tension force points downwards, complementing the forces that are trying to retard the rise of the tube. Since this mechanism depends only on the relative orientation of the twist of the structure and background field, this mechanism can be thought of as one that either (a) for a structure of fixed twist, enhances the likelihood of the rise of the structure for one particular orientation of background field, or (b) for a fixed background field, enhances the likelihood for rise of structures with a certain twist. Note that, given the limited knowledge of the field strengths in the deep interior of the Sun, for this work, we have little guidance on our choice of $B_s$ which sets the relative strengths of the background poloidal field to the axial field of the tube.  However, our experience with instability simulations \cite[e.g.][]{Vasil:Brummell:2008} suggests that rising structures are not many orders of magnitudes stronger than their originating field, providing hope that the range of $B_s$ where we find interesting dynamics is realistic.

The latter point of view more easily leads to the conclusion that this selection mechanism is commensurate with the SHHR.  In the solar context, the axial field direction and the overlaying poloidal field are not independent, and this connection (via the differential rotation), allows our mechanism, which depends on twist only, to be cast as one that depends on helicity.  When applied to the solar context, our mechanism selects the correct helicity as the preferred helicity in each hemisphere to concur with the SHHR (negative helicity in the Northern hemisphere, positive in the Southern).  This preference is independent of which half of the full 22-year solar cycle is examined, since the orientation of both the overlaying background poloidal field and the toroidal field providing the axial field of the structure flip when switching after 11 years. This mechanism therefore appears to be a reasonable candidate for the origin of the SHHR since it agrees with this major component of the rules.

Interestingly, the mechanism that we have discovered provides explanations for many of the further details of the SHHR.  In particular, our mechanism can also (i)  explain the fact that the SHHR is only a weak rule, in the sense that only 60-80\% of active regions obey the rule, and therefore 20-40\% are in violation, and, (ii) provide a plausible reason for the observed increased disparity in the rule around the period between cycles.  Indeed, expanding on the latter, our model provides a prediction for the dependence of the adherence to the main SHHR over the cycle.  These further affirmations of SHHR characteristics are gleaned from a deeper examination of the mechanism, which was the focus of much of the rest of the paper.  

To facilitate this understanding, we explored the reason behind these dynamics in detail via an analytical model of the forces acting on the magnetic structure.  The success of an attempted rise depends on the internal magnetic buoyancy and tension forces of the structure as modified by the existence of the background field, as it competes with the induced tension force in the overlying large-scale background field created by the rise through that field. The effect of the introduction of large-scale background field on the flux tube is to adjust the overall buoyancy and internal tension forces in such a way that it can either facilitate or hinder its rise. Our mathematical model confirms that the predominant adjustment comes from the internal tension force, which we analytically show to be dependent on the product of $q$ and $B_{s}$, thereby revealing again the dependence on the alignment (i.e. sign) of these two entities: having the same sign provides a positive (upward) tension force that encourages rise, whereas opposite signs create a negative tension force that counters the rise. Setting the net force balance of these three forces to zero provides a quadratic equation (\ref{eq:MarginalCase_FinalEquation}), whose roots define a regime where the selection mechanism operates.  We call this region in parameter space the Selective Rise Regime (SRR).  The region depends on the parameters $q$, $B_s$ and $H_B$ (the twist, the strength of the background field and the scale height of the background field), which we consider to be the main parameters of our studies; there are other parameters that we have fixed at canonical values.  It turns out that the dependence on the configuration of the background field via $H_B$ is weak, and therefore the SRR can be considered as mainly a region of $q-B_s$ parameter space, defined by the modulus of the roots of the characteristic quadratic equation (\ref{eq:MarginalCase_FinalEquation}).  Between the (modulus of the) two roots is the SRR, where the selection mechanism operates, and one sign of twist rises preferentially over the other.  Above the largest root, both signs of twist will rise successfully.  Below the smallest root, both twists can rise or neither twist may rise depending on the nature of the roots, but structures of both signs behave similarly.    

The analytical model SRR exhibits the general characteristics of the $q-B_s$ dependence of the simulation data, and clarifies the existence of the selection mechanism as a direct result of the asymmetry of the internal tension force in the magnetic structure.  The quantitative agreement of the analytical SRR and the selection regime found from the simulations is not perfect, mainly because estimation of the retarding effect of the tension from the overlying field is difficult, and we do not account for other retarding effects, such as drainage of the axial flux driving the buoyancy and the destruction of the trailing vortices that aid the rise.
However, the analytic model and simulation results do, in tandem, provide an explanation for why violations of the SHHR that induce ``scatter" in the observations, might be expected.  The main reason for this is that structures containing twist of either sign whose strength is above the upper bound of the SRR (for a given $B_s$) will rise successfully.  If it is expected that the twist of the magnetic structures  arises randomly in the structure generation process, then some distribution of twist might be also be expected that includes strong values of either sign whose moduli lie outside the SRR.  These values would not have the bias of the SHHR and would induce violations in the observed data.

In order to test these ideas, we carried out a MC study with the aim of synthetically reproducing the nearest equivalent of SHHR observations from our simulations.  For a given $B_s$, we computed a significant number of simulations containing multiple magnetic flux structures with randomly assigned twists (and spacing) from a reasonably wide uniform distribution.  From these, we calculated the sign and strength of the net helicity that ``emerged" near the top of our simulation box.  The relative fractional distribution of these signs (and strengths) mimic the observations that lead to the SHHR at that particular $B_s$.  We associated the chosen $B_s$ with a time in the solar cycle, with small $B_s$ representing solar minimum and large $B_s$ being related to solar maximum.  We find that for low $B_s$, the fractional distribution of the signs of helicity at emergence is statistically close to 50\%-50\%.  This fractional distribution increases as $B_s$ increases, up to around 85\%-15\% in favour of the sign expected from the SHHR for the largest $B_S$ computed.  Overall, this is in good agreement with the SHHR observations, where concurrence with the rule occurs for 60-70\% of the total active regions over the whole cycle. The violations in our simulations can be checked to confirm that they mainly come from values of the twist that lie outside the SRR, although other odd effects from structure interactions can also occur in these multiple tube simulations.

A nice quality of these MC simulations is that they provide a prediction for the temporal evolution of the agreement with the SHHR, via the fractional distribution of the signs.  The distribution  depends on $B_s$ (which we associate with epoch in the cycle) since the extent of the SRR (in $|q|$) depends on $B_s$, widening as $B_s$ increases.  This means that it is increasingly difficult to violate the SHHR (i.e. possess twist that is both below any physical cap and large enough to fall above the upper SRR boundary) as $B_s$ increases, and hence the rule is obeyed more completely at higher $B_s$ (note that this assumes that other effects at very small twist are less important).  With SHHR observational data collected over more solar cycles, this prediction could be tested.

Various other models \citep{Longcope:Fisher:Pevtsov:1998, Choudhuri:2003} have attempted to address the issue of the origin of SHHR, as described in the introduction, and here it is pertinent to describe the main differences in the results from our model and the others, and what each can contribute to our understanding.
A major distinction between our model and most others is that the magnetic flux structures in most other models are initially untwisted and then acquire the appropriate twist by some mechanism during transit, whereas in our model, all flux structures possess some twist initially (of random size and sign) and then the structures with the ``correct'' twist are selected for rise by a filtering mechanism.
The models acquiring twist either do it directly, by rising into and then merging with other field (e.g. \citet{Choudhuri:2003}), or indirectly, by acquiring writhe from dynamics influenced by the Coriolis force which then, by helicity conservation, induces a compensatory twist of the ``correct" sign (e.g. LFP).  A big difference between the models is then the degree to which they explain the scatter in the adherence to the SHHR.  For example, the Choudhuri model, as presented, predicts 100\% agreement with the SHHR, except perhaps at solar minimum.  The $\Sigma$ mechanism of LFP, however, does produce a scatter since the twist is generated (indirectly) from interaction with the convective turbulence, and therefore a distribution of twist will result.  It is worth noting that most of these previous models assume that the magnetic flux structures have zero twist initially, which seems unlikely. Allowing an initial distribution of twists (as in our model) would actually provide another source of scatter in the resultant twist observations in these models too.

The different models also provide different results for the solar cycle dependence of the SHHR, in particular, with respect to the cycle dependence of the scatter of the observations and the anti-SHHR behaviour observed at solar minimum \citep{Seehafer:1990, Pevtsov:Canfield:Metcalf:1995, Abramenko:Wang:Yurchishin:1997, Bao:Zhang:1998, Bao:Ai:Zhang:2000, Bao:Pevtsov:Wang:Zhang:2000, Hagino:Sakurai:2005, Hao:Zhang:2011}.  For example, our model predicts that agreement with the SHHR increases in the rise to solar maximum and decreases towards solar minimum, whereas the model of LFP predicts no variation in the agreement over the solar cycle.  Our model and Choudhuri's model both predict anti-SHHR behaviour in the transition between cycles if the switch of polarity of the poloidal and toroidal components of the dynamo field are slightly out of phase.  The model of LFP does not know about this phase information and depends only on the influence of rotation on convection which is cycle independent.

All these mechanisms are complementary in some sense, and could all be contributing to the ultimate helicity content of rising structures and therefore the SHHR observations.  One can imagine a scenario where our model is dominant near the base of the convection zone, LFP's model operates in the bulk of the convection zone, and Choudhuri's model contributes near the surface.
Our model is a very complete one, in the sense that it individually not only explains the preferred helicity of active regions, as required by the SHHR, but also the scatter in the observations, and possibly other elements, such as the anti-SHHR behaviour at the beginning of the cycle.  We have already begun full 3D simulations to confirm our mechanism in that context.  In future work, we will expand these to simulations that include a rotationally-influenced convective zone through which the structure must transit. In this manner, we will be able to examine the contributions from the various mechanisms, whilst moving away from the concept of isolated (and thin) flux tubes, as clearly the dynamics of magnetic flux concentrations are undeniably different.

One extra highly conjectural but fascinating possibility from our model is that it could potentially offer an indirect understanding of the elusive magnetic nature of the solar interior.  Since the degree of agreement with the SHHR (the scatter fraction) is significantly dependent on the strength of the interior field ($B_s)$ in our model (but only weakly dependent on the configuration, $H_B$), the exact observed fractional degree of  scatter might provide a probe into the interior field strength at least, if not the vertical variation of the field.  We therefore look forward to improved multi-cycle observations of the SHHR.

We thank Dongwook Lee for significant expert help with initially setting up the model in the FLASH code. The authors acknowledge funding from NSF AST grant 1908010  and NASA DRIVE Center Phase 1 grant 80NSSC20K0602 (subcontract to University of California, Santa Cruz). The authors also acknowledge NSF XSEDE grants for access to the Texas Advanced Computing Center (TACC \url{http://www.tacc.utexas.edu}) at The University of Texas at Austin for providing HPC, visualization, and database resources that have contributed to the research results reported within this paper. We further acknowledge use of the Lux and Hyades supercomputers at University of California Santa Cruz, that were funded by NSF MRI grants AST 1828315 and 1229745 respectively. We thank the anonymous referee for helpful suggestions.

\appendix

The Appendices here contain material related to the robustness of our results. We have tested robustness to different models for the thermal initial conditions, different models of the magnetic configuration of the flux structure, different models for the background field, changing the boundary conditions, and varying the (dimensionless) magnetic diffusivity.  The ultimate conclusion is that the results are very robust, and therefore much of this work is withdrawn from the main text in order to highlight the more salient points.  Here, we outline the first two of these robustness checks because they are pertinent to what is presented in the paper.  The initial thermal conditions are relevant to understanding the forces in the model in Section \ref{sec:ModelsMethods}, and the alternative model for the structure of the tube's internal fields is used in Section \ref{sec:PreviousResults} (both to actively demonstrate robustness and to provide different results from the previous letter).

\bigskip
\section{Different initial thermodynamic conditions} \label{Appendix_InitialThermodynamicConditions}

The effect of the introduction of a magnetic flux structure into a stratified gas is to add an extra magnetic pressure to the total pressure. However, it is generally assumed that the total pressure equilibrates quickly under the circumstances of interest, thereby adjusting the gas pressure in the magnetic concentration. This change in gas pressure can lead to a change in density, temperature, or both. There is no unique way of specifying the initial thermodynamic conditions inside the magnetic flux tube. In order to check robustness of our results to these different formulations, we have tested two commonly-used sets of initial thermal conditions.  The crucial result to highlight here is that changing these makes essentially no difference to our major results and only very minor differences to any of the dynamics of the problem.  We supply some description here to verify this fact, and to provide details of the formulations that we use in the main text (Case 2).  Incidentally, we did also examine a case that did not initial obey the condition of total pressure equilibration, which would be considered unphysical by most.  Since this case again produced very similar results, it is not included here.

When a flux tube is introduced into the stratification,  the assumption of total pressure balance introduced above dictates that
\begin{equation} \label{eq:TotalPressureBalance}
    p_{gas,in} + p_{B,in} = p_{gas,out} + p_{B,out} 
\end{equation}
throughout the interior or at least at the edge of the tube.
Here, $p_{gas,in}$ and $p_{B,in}$ are the gas and magnetic pressure inside the flux tube respectively, and,  similarly, $p_{gas,out}$ and $p_{B,out}$ are the gas and magnetic pressure outside the flux tube.   The latter two are known, given respectively by Equation \ref{eq:PolytropicModel} and 
\begin{equation} \label{eq:OutsideMagneticPressure}
p_{B,out} = \dfrac{|\mathbf{B_{out}}|^{2}}{2} = B_{s}^{2}~\text{exp} \Big(\dfrac{y_{c}-y}{H_{B}} \Big).
\end{equation}
These therefore depend solely on height $y$.
If $B_{s} \neq 0$, the magnetic pressure inside the flux tube, $p_{B,in}$, depends on the interior magnetic field $\mathbf{B_{in}}$ which includes contributions from both the magnetic field of the flux tube ($\mathbf{B_{tube}}$) and the background field ($\mathbf{B_{back}}$), and so, from Equation (\ref{eq:TopHatProfile_B})

\begin{equation} \label{eq:EffectiveMagneticPressure}
    p_{B,in} = \dfrac{|\mathbf{B_{in}}|^{2}}{2} = \dfrac{1+4q^{2}r^{2}-4q(y-y_{c})B_{back}+B_{back}^{2}}{2}
\end{equation}
where we have used $x^{2}+(y-y_{c})^{2} = r^{2}$ for simplicity.  This is a function of $x$ and $y$.  

Equations (\ref{eq:TotalPressureBalance}, \ref{eq:EffectiveMagneticPressure} and \ref{eq:OutsideMagneticPressure}) together dictate an adjusted gas pressure inside the concentration, $p_{gas,in}$.  In forming a complete set of initial conditions, it therefore remains to determine how to compensate for this adjustment with the temperature and the density inside the flux structure. We consider two cases.

\begin{figure}
    \centering
    \includegraphics[width=\textwidth,height=10cm]{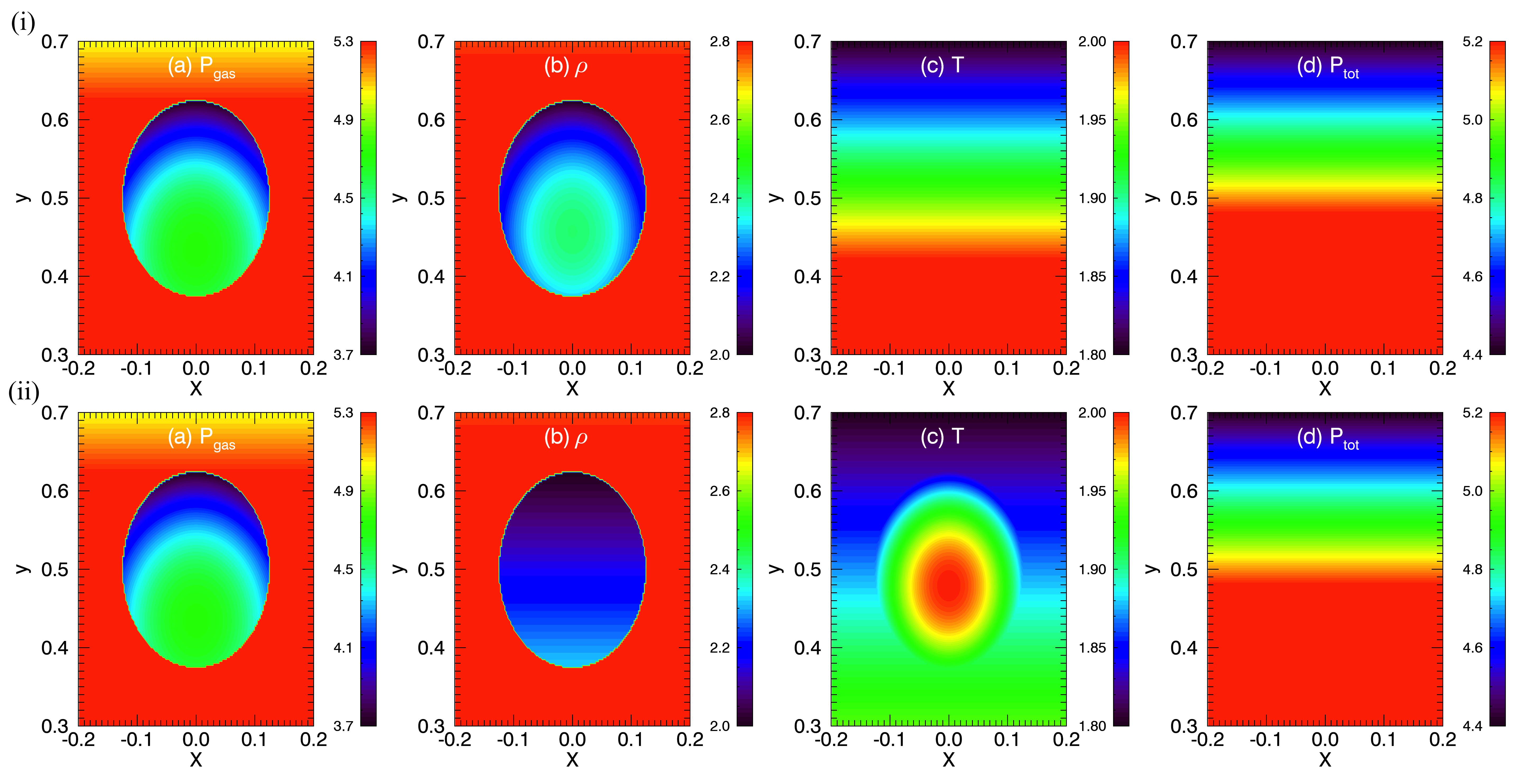}
    \caption{Thermodynamic initial conditions for (i) Case 1 and (ii) Case 2:  (a) gas pressure, (b) density, (c) temperature, and (d) total pressure. A subset of the domain is shown for clarity.}
    \label{fig:C1_ThermodynamicConditions}
\end{figure}

\noindent
\underbar{Case 1:}~~The
simplest setup is to assume that both the temperature also equilibrates quickly to match the external thermodynamics. The assumptions of total pressure and temperature equilibration are perhaps the most commonly used ones to explain the existence of magnetic buoyancy in magnetic field concentrations. Under these assumptions, we have

\begin{subequations}
 \begin{equation}
     p_{gas,in}(x,y) = p_{gas,out}(y) + p_{B,out}(y) - p_{B,in}(x,y),
 \end{equation}
 \begin{equation}
     T_{gas,in}(y) = T_{gas,out}(y),
 \end{equation}
 \begin{equation}
     \rho_{gas,in}(x,y) = \dfrac{p_{gas,in}(x,y)}{T_{gas,in}(y)},
 \end{equation}
\end{subequations}
\noindent
Above, $T_{gas,in}$, the gas temperature inside the concentration, gets set to the polytropic gas temperature outside, $T_{gas,out}$, by the assumption, and then the density inside the concentration is determined by the ideal gas equation from this temperature and the adjusted gas pressure $p_{gas,in}$.
Figure (\ref{fig:C1_ThermodynamicConditions}i) shows the properties of these initial conditions (zoomed in to focus on the flux tube and its near exterior). 

\noindent
\underbar{Case 2:}~~A somewhat less restrictive interpretation of the internal thermodynamics was used by some authors, \citep[e.g.][]{Hughes:Falle:Joarder:1998}. Here, the temperature is considered merely continuous at the edge of the tube and allowed to vary such that the density and total pressure inside the tube are solely a function of height. To achieve this, the pressure and temperature at the edge of the concentration ($r=\sqrt{x^2+(y-y_c)^2}=R$) are used to set the adjusted density at the edge:

\begin{subequations}
\label{eq:ThermodynamicConditionEdges}
 \begin{equation}
     p_{gas,edge}(y) = p_{gas,out}(y) + p_{B,out}(y) - p_{B,in}(x,y)|_{r=R},
 \end{equation}
 \begin{equation}
     T_{gas,edge}(y) = T_{gas,out}(y),
 \end{equation}
 \begin{equation}
     \rho_{gas,edge}(y) = \dfrac{p_{gas,edge}(y)}{T_{gas,edge}(y)},
 \end{equation}
\end{subequations}
This density is then propagated  over all $x$, and, with the adjusted internal pressure from Equation (\ref{eq:TotalPressureBalance}), used to create the internal temperature:

\begin{subequations}
 \label{eq:Case2_Thermodynamic}
\begin{equation}
p_{gas,in}(x,y) = p_{gas,out}(y) + p_{B,out}(y) - p_{B,in}(x,y)
\end{equation}
\begin{equation}
\rho_{gas,in}(y) = \rho_{gas,edge}(y)
\end{equation}
\begin{equation}
T_{gas,in}(x,y) = \dfrac{p_{gas,in}(x,y)}{\rho_{gas,in}(y)}
\end{equation}
\end{subequations}
Figure (\ref{fig:C1_ThermodynamicConditions}ii) shows the properties for Case 2.

\begin{figure}
	\includegraphics[width=\columnwidth, height=10 cm]{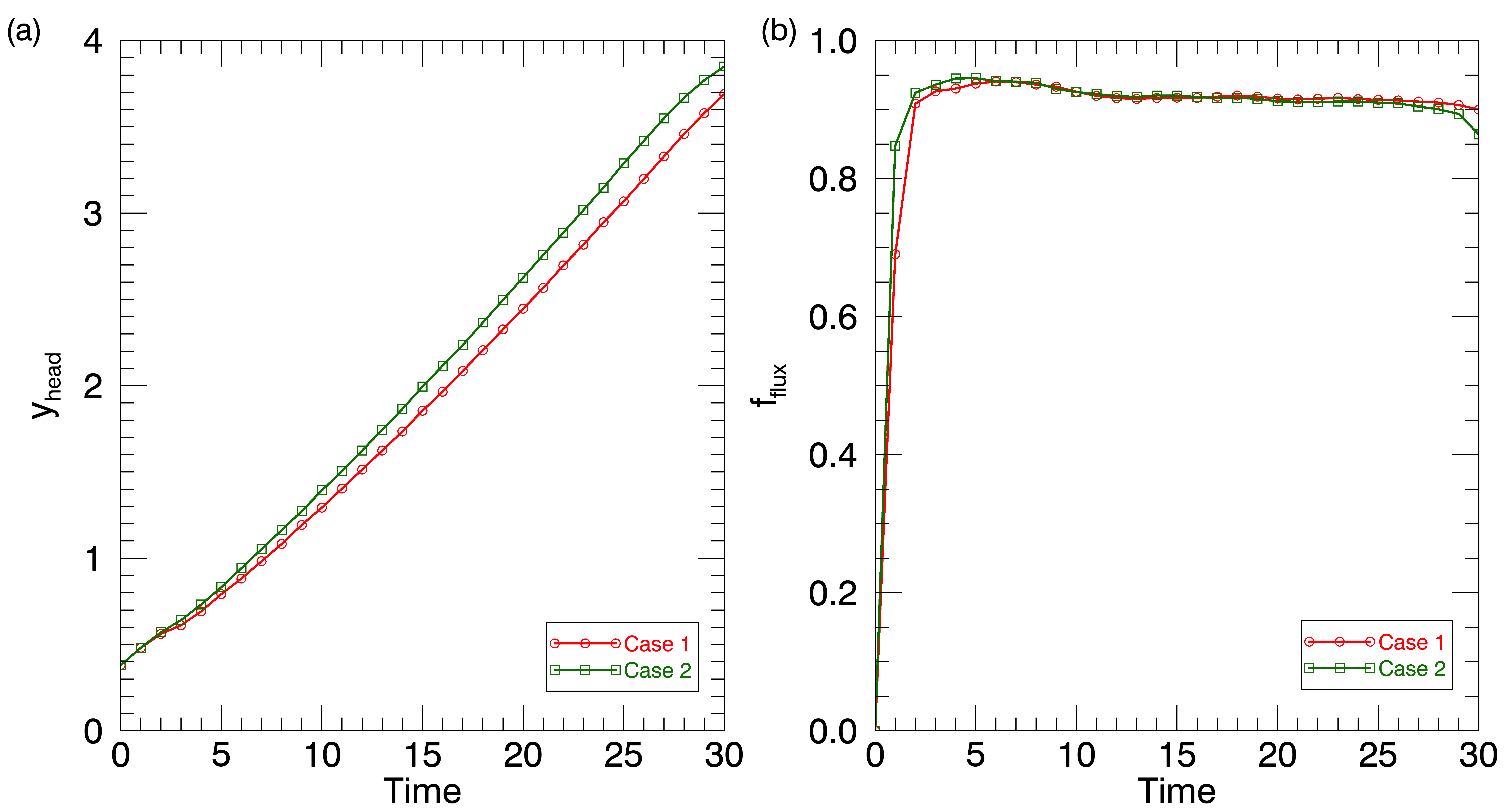}
    \caption{Comparison of (a) vertical position of the head of the rising flux tube and (b) rising flux fractions for three different initial thermodynamic conditions for $B_{s}=0$. Flux fraction (calculated with $v_{y,threshold}=0.03$) in (b) has been curtailed at the time when the flux tube structure reaches the top of the simulation box.}
    \label{fig:C1_C2_C3_Compare}
\end{figure}

Figure \ref{fig:C1_C2_C3_Compare} compares the vertical position of the head of the magnetic flux structure and rising flux fraction from Equation (\ref{eq:flux_fraction_definition}) for both cases for simulations in the absence of background field ($B_s=0$). We find the location of the head of the flux tube by measuring the vertical distance to the maximum of  $B_{z}$ on the central plane, $x=0$, as a function of time.  We curtail this plot at the time (denoted by the rise time) when the flux tube structure has reached the top of the simulation box.  The two cases have quite different internal density functions inside the initial magnetic structure, since, from above, in Case 1, is a function of $x$ and $y$  whereas for Case 2 it is solely a function of $y$. Despite these differences in the initial thermodynamic conditions, the time it takes for the flux tube to reach the top of the simulation box is very similar (see figure \ref{fig:C1_C2_C3_Compare}a), with Case 1 being slightly slower since its integrated density perturbation is slightly smaller. The rising flux fractions, as shown in figure (\ref{fig:C1_C2_C3_Compare}b), are also very similar, showing the usual characteristics of successful rise. Hence, it appears that the simulation results are robust to different interpretations of the initial thermodynamic conditions. We have used Case 2 as our canonical case for all the results of this paper.

\section{Initial axial magnetic field profile}
As mentioned and shown earlier in Section 1, we have also tested the effect of switching the flux tube profile from the top hat profile of Paper I to a Gaussian profile, as has been used by various other authors \citep[e.g.][]{Cheung:Moreno:Schussler:2006}. The top-hat profile is given simply by 

\begin{equation}
\label{eq:TopHatProfile}
    B_{z}(r,\theta) = 
    \begin{cases}
    1 & \text{if}~  r \le R_{o}\\
    0 & \text{if}~  r > R_{o},       
    \end{cases}
\end{equation}
whereas a Gaussian profile is given by
\begin{equation}
 \label{eq:GaussianProfile}
    B_{z}(r,\theta) = 
    \begin{cases}
    B_{o}e^{-r^{2}/R^{2}} & \text{if}~  r \le 2R_{o}\\
    0 & \text{if}~  r > 2R_{o}.        
    \end{cases}
\end{equation}
We choose $B_{o}$ such that the integral of $B_z^2$ over the concentration is the same in both cases, and therefore the buoyancy forces induced by each are also the same.

\begin{figure}
	\includegraphics[width=\columnwidth, height=10cm]{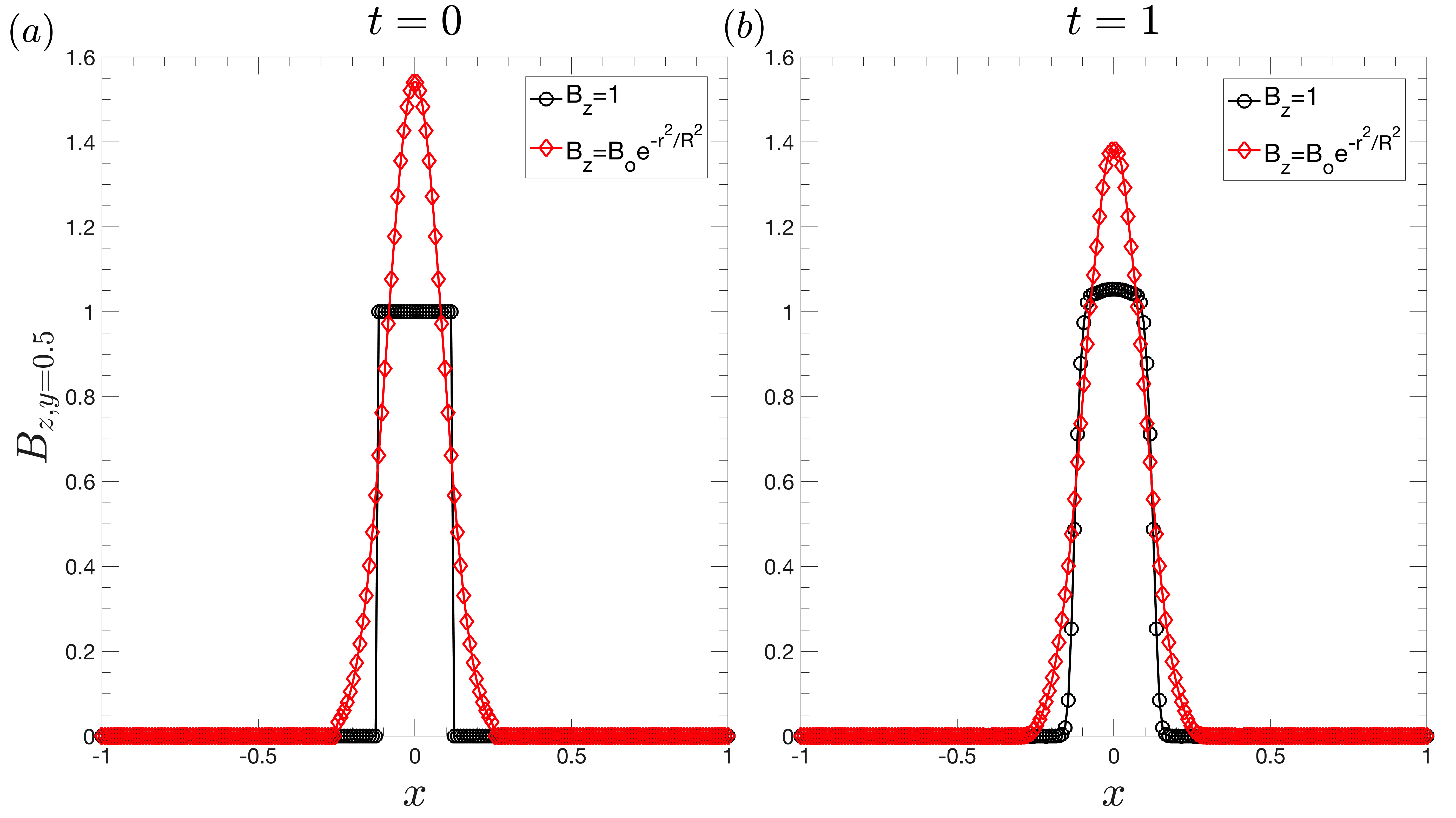}
    \caption{Line cuts of $B_{z}$ against $x$ at the height of the center of the magnetic concentration, $y=0.5$, for simulations started with both top hat and Gaussian  profiles inside the magnetic structure. (a) The initial condition at $t=0$.  (b) A short time later, at $t=1$}
    \label{fig:MagneticIC_figure}
\end{figure}

Fig.~\ref{fig:MagneticIC_figure} shows line cuts of these two initial magnetic field profiles through the center of the flux tube at a height of $y=0.5$.  A top hat profile has unrealistically strong magnetic field gradients at the edge of the flux tube, although this appears not to be an issue, since, at the parameters of the simulations here, diffusion acts quickly to smooth these gradients. This can be seen in  Fig.~\ref{fig:MagneticIC_figure}b, which shows the profiles at $t=1$.  Even at this very early stage in the rise, the top hat profile has been smoothed to a more physical profiles and the two profiles have become very similar.  

\begin{figure}
	\includegraphics[width=\columnwidth, height=9cm]{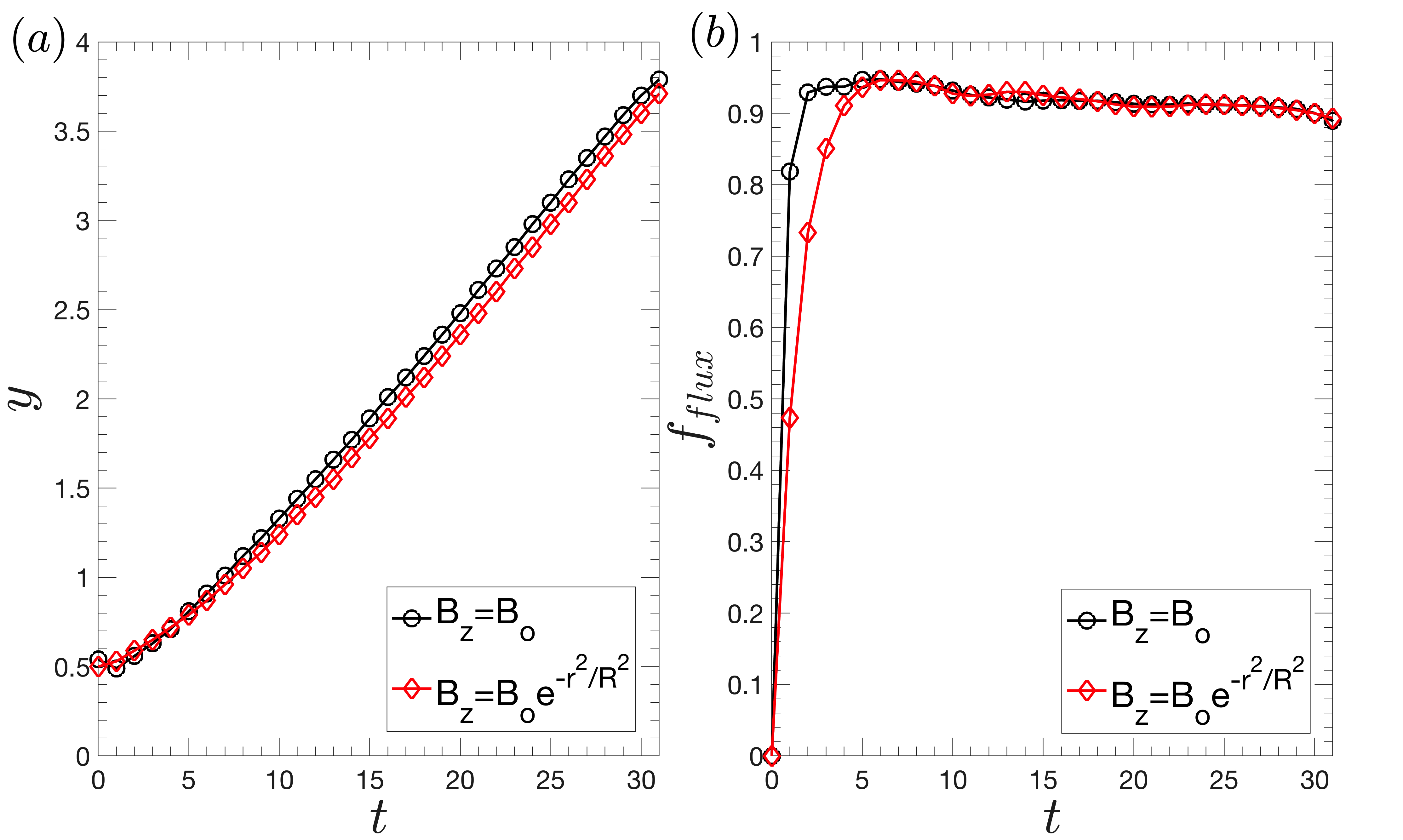}
    \caption{(a) Rise time and (b) flux fraction for top-hat and Gaussian magnetic field profiles at $B_{s}=0$}
    \label{fig:TopHatGaussian_RT_FF_figure}
\end{figure}

Ultimately, we again found that the choice of profile had little consequences for the dynamics of interest. Fig.~\ref{fig:TopHatGaussian_RT_FF_figure} shows the rise time and flux fraction as a function of time for both these case (when $B_{s}=0$) and it can be seen that the dynamics are very similar. 
A more stringent test is the comparison of, for example, Fig.~\ref{fig:Figure2_Paper2} in this paper with Fig.~2 of Paper I. From these, it is clear that the results are not substantially influenced by the choice of profile.

\bibliography{Paper_Manek_Brummell}{}
\bibliographystyle{aasjournal}

\end{document}